\documentclass[pra,reprint,twocolumn,superscriptaddress,showpacs,floatfix]{revtex4-1}
\usepackage{amsmath}
\usepackage{mathrsfs}
\usepackage{txfonts}
\usepackage{amssymb}
\usepackage{graphicx}
\usepackage{hyperref}
\usepackage{ulem}
\usepackage{xcolor} 
\usepackage{color}
 \usepackage{overpic}
 \usepackage{psfrag}
\usepackage{tabularx}
\usepackage{multirow}
\usepackage{array}
\usepackage{placeins}
\newcommand{\PreserveBackslash}[1]{\let\temp=\\#1\let\\=\temp}

\usepackage{bm}
\usepackage{dsfont}
\usepackage{mathtools}

\makeatletter

\begin{document}

\title{Entanglement entropy for the one-dimensional flat-band ferromagnetic Tasaki model:\\  spontaneous symmetry breaking with one type-B Goldstone mode}

\author{Huan-Qiang Zhou}
\affiliation{Centre for Modern Physics, Chongqing University, Chongqing 400044, The People's Republic of China}

\author{Qian-Qian Shi}
\affiliation{Centre for Modern Physics, Chongqing University, Chongqing 400044, The People's Republic of China}

\author{Ian P. McCulloch}
\affiliation{Department of Physics, National Tsing Hua University, Hsinchu 30013, Taiwan}
\affiliation{Frontier Center for Theory and Computation, National Tsing Hua University, Hsinchu 30013, Taiwan}

\author{J. O. Fj{\ae}restad}
\affiliation{ Center for Quantum Spintronics, Department of Physics,
	Norwegian University of Science and Technology, NO-7491 Trondheim, Norway}

\begin{abstract}
The one-dimensional flat-band ferromagnetic Tasaki model exhibits spontaneous symmetry breaking  from ${\rm SU}(2)$ to ${\rm U}(1)$ with one type-B Goldstone mode, featuring that the highest weight state is entangled at quarter filling, but there is always a choice to keep the highest weight state unentangled  away from quarter filling. It is found that the ground-state degeneracies under both periodic and open boundary conditions constitute essentially the Fibonacci sequences, behaving asymptotically as the golden spiral - a self-similar geometric object. A set of orthonormal basis states  are generated from the repeated action of the lowering operator of the symmetry group ${\rm SU}(2)$ on the highest weight state at a specific filling.  In particular, it is possible to construct the orthonormal basis states reflecting an abstract fractal underlying the ground-state subspace, which are permutation-invariant away from quarter filling, but not at quarter filling. As a consequence, there exists a singularity that accounts for the emergence of the saturated flat-band ferromagnetism at quarter filling. We perform a systematic finite system-size scaling analysis of the entanglement entropy,  thus confirming that it scales logarithmically with the block size in the thermodynamic limit, with the prefactor being half the number of type-B Goldstone modes, for the orthonormal basis states  at and away from quarter filling. 
\end{abstract}
\maketitle

\section{Introduction}

The origin of ferromagnetism continues to be a challenging problem in strongly correlated itinerant electron systems, due to the fact that 
the intricate competition between the wave-like nature and the particle-like nature of electrons is generically not analytically tractable. However, a few mathematically rigorous results on ferromagnetism are available~\cite{tasakibook}. For instance,  an early result for the Hubbard model, introduced as a tight-binding minimal model, was due to Nagaoka~\cite{nagaoka}. This is consistent with the so-called Stoner's criterion, when the on-site repulsive Coulomb interaction is infinitely large, if only one hole is present. Another direction follows from the criterion when the density of states at the Fermi level is divergent, which appears to be more promising in understanding the emergence of ferromagnetism. This phenomenon is known as the flat-band ferromagnetism~\cite{mielke,tasaki1, moreflatband} that is described by the variants of the Hubbard model in a variety of decorated lattices, featuring a flat band at the bottom of the single-electron spectrum, subject to a macroscopic degeneracy. 
A model with a flat band may be systematically constructed in terms of the line graphs by Mielke~\cite{mielke} and the cell constructions by Tasaki~\cite{tasaki1}, and exhibit the saturated ferromagnetism when the electron number is the same as the number of the lattice unit cells, which is referred to as quarter filling in one dimension.

As is well-known, the quantum spin-$1/2$ ${\rm SU}(2)$ ferromagnetic Heisenberg model constitutes a paradigmatic example for spontaneous symmetry breaking (SSB) with type-B Goldstone modes (GMs).  Indeed, it appears to be necessary to make   
a distinction between type-A and type-B GMs to understand the physics behind quantum many-body systems undergoing SSB, as follows from the counting rule of GMs~\cite{watanabe}.
Actually, not until quite recently has significant progress been made in a proper classification of SSB with distinct types of GMs, as a result of a long-term pursuit~\cite{goldstone,Nambu,nielsen,schafer, miransky, nambu, nicolis, brauner1, watanabe, NG}. In fact, many fascinating physical phenomena  in condensed matter physics, e.g., antiferromagnetic order and superfluidity, are attributed to SSB with type-A GMs. In contrast, it is not so common to encounter SSB with type-B GMs in quantum many-body systems. However,  a few exceptional examples in the guise of quantum many-body spin systems have been exposed to exhibit SSB with type-B GMs~\cite{FMGM,LLspin1,Golden, finitesize, stsu4, dtmodel}, which always involve a variety of ferromagnetic interactions. A notable feature for quantum many-body systems undergoing SSB with type-B GMs is that they are described by the so-called frustration-free Hamiltonians~\cite{tasakibook}.  As a result,   highly degenerate ground states arising from SSB with type-B GMs are exactly solvable. Indeed,  the orthonormal basis states, which span the ground-state subspace, are generated from the repeated action of the lowering operator(s) on the highest weight state and admit an exact Schmidt decomposition, which in turn implies that the orthonormal basis states exhibit  self-similarities in the real space~\cite{FMGM}. Here, we remark that the entanglement entropy for the spin-$1/2$ ${\rm SU}(2)$ ferromagnetic Heisenberg model has been investigated in Refs.~\cite{popkov,doyon,ding,hqzhou}. In fact, it has been found that  the entanglement entropy scales logarithmically with the block size in the thermodynamic limit, with the prefactor being half the number of type-B GMs for the orthonormal basis states~\cite{FMGM,LLspin1,Golden, finitesize, stsu4, dtmodel}.
Given that the effective spin model for the  flat-band ferromagnetic Tasaki model at quarter filling is the spin-$1/2$ ${\rm SU}(2)$ ferromagnetic Heisenberg model  when the on-site repulsive Coulomb interaction is sufficiently strong~\cite{penc}, one may anticipate that 
SSB with type-B GMs occurs in the strongly correlated itinerant electron systems exhibiting the flat-band ferromagnetism. 
Hence, it is highly desirable to clarify  whether or not the entanglement entropy for the flat-band ferromagnetic Tasaki model behaves in the 
same way as that for quantum many-body spin systems undergoing SSB with type-B GMs.

In this work, we focus on the one-dimensional version of the flat-band ferromagnetic Tasaki model. As it turns out, it exhibits SSB from ${\rm SU}(2)$ to ${\rm U}(1)$ with one type-B GM, featuring that the highest weight state is entangled at quarter filling but not necessarily entangled away from quarter filling, in sharp contrast to  the spin-$1/2$ ${\rm SU}(2)$ ferromagnetic Heisenberg model. It is found that the ground-state degeneracies under both periodic  boundary conditions (PBCs) and open boundary conditions (OBCs) constitute essentially the Fibonacci sequences. As a result, they grow exponentially with the system size, thus behaving asymptotically as the golden spiral - a self-similar geometric object. Hence, the residual entropy per the lattice unit cell is non-zero in the grand canonical ensemble. Indeed, it is related to the golden ratio, as already noticed in Ref.~\cite{goldenratio}.

At quarter filling, the orthonormal basis states, generated from the repeated action of the lowering operator of the symmetry group ${\rm SU}(2)$ on the entangled highest weight state, are not permutation-invariant with respect to the lattice unit cells, consisting of the two nearest-neighbor lattice sites. As a consequence, if the system is partitioned into a block and its environment, then the entanglement entropy does depend on what types of  boundary conditions, i.e., PBCs or OBCs, are adopted. In contrast,  the entanglement entropy does not depend on what types of boundary conditions are adopted away from quarter filling, if one chooses the highest weight state at a specific filling properly such that  degenerate ground states generated from the repeated action of the lowering operator of the symmetry group ${\rm SU}(2)$ on this highest weight state are permutation-invariant with respect to an emergent unit cell, different from the lattice unit cell.
Interestingly, the presence of  such emergent unit cells yields a dense subset of the values of the filling that covers the entire range  in the thermodynamic limit, thus ensuring the existence of an intrinsic abstract fractal~\cite{hqzhou}, as already reflected in the fact that the ground-state degeneracies under PBCs  and OBCs behave asymptotically as the self-similar golden spiral. 

In addition, an exact graded matrix product state (MPS) representation is constructed for the flat-band ferromagnetic Tasaki model, which provides an alternative powerful means to account for the emergence of the nearly-flat-band ferromagnetism, with the frustration-free nature of the perturbed Hamiltonian being left intact~\cite{nearlyflat,nearlyflat1}. Indeed, degenerate ground states remain to be ferromagnetic, as long as the  on-site repulsive Coulomb interaction and the band gap are sufficiently large, though the flat band becomes a nearly-flat band that is now dispersive. Generically, this  representation opens up the possibility for both analytical and numerical approaches to the nearly-flat-band ferromagnetism in this type of strongly correlated itinerant electron models in the context of a  graded tensor network  representation~\cite{gradpeps}, including a graded matrix product state (MPS) representation and a graded projected  entangled-pair state (PEPS) representation, first introduced in a fermionic realization~\cite{kraus} (see also~\cite{fermionpeps,fermionpeps1}).

In particular,  a finite system-size scaling analysis is performed for the entanglement entropy~\cite{finitesize}. Here we stress that the exact graded MPS representations for degenerate ground states may be exploited  as a powerful and efficient means to facilitate the evaluation of many physical observables,  including the entanglement entropy. It is found that the entanglement entropy scales logarithmically with the block size in the thermodynamic limit, with the prefactor being half the number of type-B GMs $N_B$ for  the orthonormal basis states at and away from quarter filling. As such, the flat-band ferromagnetic Tasaki model behaves very much in the same way as quantum many-body spin systems undergoing SSB with type-B GMs, as far as the scaling behaviors of the entanglement entropy are concerned. However,  some extra complications arise from the nature of strongly correlated itinerant electrons for the flat-band ferromagnetic Tasaki model, since the number of electrons is conserved, in addition to the symmetry group $\rm{SU(2)}$ in the spin sector. A remarkable feature  concerns the distinct behaviors of the entanglement entropy at and away from quarter filling, thus revealing a singularity that itself may be exploited to account for why the saturated flat-band ferromagnetism {\it only} emerges at quarter filling~\cite{tasakibook,tasaki1}.

The layout of this paper is  as follows. In Section~\ref{tasakimodel}, we briefly recall the flat-band ferromagnetic Tasaki model on a decorated lattice.
In Section~\ref{obs},  the orthonormal basis states are constructed at and away from quarter filling. In Section~\ref{gsd},  the ground-state degeneracies under PBCs and OBCs are discussed.  In Section~\ref{esd}, an exact Schmidt  decomposition is exposed  for the orthonormal basis states constructed from a specific choice of the highest weight state away from quarter filling, thus revealing their self-similarities in the real space, but this feature is lost at quarter filling. In Section~\ref{gmps}, an exact graded MPS representation for highly degenerate ground states are presented, which are orthonormal basis states  constructed from the (periodic) highest weight state at a specific filling, including quarter, $1/8$ and  $1/12$ fillings. In Section~\ref{ee}, we perform a finite system-size scaling analysis of the entanglement entropy at and away from quarter filling. The last Section is devoted to concluding remarks.

\section{The flat-band ferromagnetic Tasaki model on a decorated lattice}~\label{tasakimodel}

Let us set the notations to define the flat-band ferromagnetic Tasaki model on a $d$-dimensional decorated hypercubic lattice.
Following Tasaki~\cite{tasakibook}, $\mathscr{E}$ denotes the set of sites in the lattice with unit lattice
spacing.  Introduce a new site in the middle of each bond of the lattice, with the collection of all such
sites being denoted by $\mathscr{I}$. Then the decorated hypercubic lattice  $\Lambda$ is the union of $\mathscr{E}$ and  $\mathscr{I}$, i.e.,
$\Lambda=\mathscr{I} \cup \mathscr{E}$. We stress that both $\mathscr{E}$ and  $\mathscr{I}$ are identical under PBCs
and OBCs, with the {\it only} difference being that there are no bonds between the lattice sites at the edges under OBCs.

The flat-band ferromagnetic Tasaki  model is a variant of the Hubbard model on the decorated 
lattice $\Lambda$, and is described by the Hamiltonian~\cite{tasaki1}
\begin{equation}
	\hat{H}= t\sum_{u\in \mathscr{I},\sigma=\uparrow, \downarrow}{\hat b}_{u,\sigma}^\dagger{\hat b}_{u,\sigma}+
	U\sum_{x\in \Lambda}{\hat n}_{x,\uparrow}{\hat n}_{x,\downarrow}, \label{ham}
\end{equation}
where  ${\hat b}_{u,\sigma}={\hat c}_{u,\sigma}+\nu\sum_{p\in \mathscr{E},(|p-u|=1/2)}{\hat c}_{p,\sigma}$ and  ${\hat b}^\dagger_{u,\sigma}$ is the Hermitian conjugate for $u\in \mathscr{I}$ and ${\hat n}_{x,\sigma} ={\hat c}^\dagger_{x,\sigma} {\hat c}_{x,\sigma}$ for $x\in \Lambda$, with ${\hat c}_{x,\sigma}$ and ${\hat c}^\dagger_{x,\sigma}$ being the destruction and creation operators of electrons for spin $\sigma$ ($\sigma =\uparrow$ and $\downarrow $) at the lattice site $x$, respectively. In addition, $t$  is a hopping  parameter, $U$ describes the on-site repulsive Coulomb interaction, and $\nu$ is a real number. From now on, we assume that $t>0, U> 0$ and  $\nu>0$. We emphasize that in the above definition of ${\hat b}_{u,\sigma}$, the sum over $p$, subject to the constraints $|p-u|=1/2$,  depends on what types of boundary conditions are adopted.

The Hamiltonian (\ref{ham}) possesses the symmetry group  $\rm{U(1)} \times\rm{SU(2)}$, with  $\rm{U(1)}$ in the charge sector being generated by ${\hat N}= \sum _{x\in \Lambda} ({\hat n}_{x,\uparrow}+{\hat n}_{x,\downarrow})$
and  $\rm{SU(2)}$  in the spin sector being generated by the three generators:  ${\hat S}^+=\sum_{x\in \Lambda} {\hat S}^+_x$, ${\hat S}^-=\sum_{x\in \Lambda}{\hat S}^-_x$ and ${\hat S}^z=\sum_{x\in \Lambda}{\hat S}^z_x$, satisfying $[{\hat S}^z,{\hat S}^+]={\hat S}^+$, $[{\hat S}^z,{\hat S}^-]=-{\hat S}^-$ and $[{\hat S}^+,{\hat S}^-]=2{\hat S}^z$, with ${\hat S}^z_x$, ${\hat S}^+_x$ and ${\hat S}^-_x$ being defined by ${\hat S}^z_x=({\hat n}_{x \uparrow}-{\hat n}_{x \downarrow})/2$, $ {\hat S}^+_x={\hat c}_{x\uparrow}^\dagger {\hat c}_{x\downarrow}$, and $ {\hat S}^-_x={\hat c}_{x\downarrow}^\dagger {\hat c}_{x\uparrow}$.
For ${\hat S}^-$ and ${\hat S}^+$, one may choose ${\hat S}^+_x$ and ${\hat S}^-_x$ as the interpolating fields~\cite{nielsen, inter}.
Given that $\langle[{\hat S}^-,{\hat S}^+_x]\rangle\propto \langle {\hat S}^z_x\rangle$ and $\langle[{\hat S}^+, {\hat S}^-_x]\rangle\propto \langle {\hat S}^z_x\rangle$, we are led to conclude that
the two generators ${\hat S}^-$ and ${\hat S}^+$ are spontaneously broken, with $\langle {\hat S}^z_x\rangle$  being the local order parameter. 
Here, the expectation value has been taken with respect to the highest weight state, which in turn depends on the filling fraction $f$ and the dimension $d$, where the filling fraction $f$ is defined as $f=N/|\Lambda|$, corresponding to the filling $f/2$. An explicit expression for the highest weight state $|\psi_0\rangle$ at a specific filling always yields non-zero local order parameter $\langle {\hat S}^z_x\rangle$, as discussed below when $d=1$.  
Hence, we are led to conclude that the model (\ref{ham}) undergoes SSB from $\rm{SU(2)}$ to  ${\rm U(1)}$, with the number of type-B GMs being one. 
To the best of our knowledge,  although many quantum many-body spin systems are revealed to exhibit SSB with type-B GMs, the flat-band ferromagnetic Tasaki  model is the {\it only} known model that exhibits SSB with type-B GMs as an itinerant electron system.

From now on, we restrict ourselves to the one-dimensional version of the Tasaki model on a decorated lattice $\Lambda$, with the lattice unit cells consisting of two nearest-neighbor lattice sites.  We stress that the filling fraction $f$ defined here is with respect to the lattice sites, instead of the lattice unit cells consisting of the two nearest-neighbor lattice sites, in order to keep consistency with the standard convention. For brevity, we introduce $L$ to denote the system size $|\Lambda|$. Hence, the lattice sites in $\mathscr{E}$ are labeled as $\{ 1,2,\ldots,L/2\}$ and the lattice sites in $\mathscr{I}$ are labeled as $\{ 3/2,5/2,\ldots,(L+1)/2\}$, with $L$ being always even. When the electron number is the same as the number of the lattice unit cells, the filling fraction $f$ is equal to $1/2$, which has been referred to as quarter filling.
At quarter filling, the highest weight state $|\psi_0\rangle$ under PBCs takes the form 
\begin{equation}
|\psi_0\rangle= \frac {1} {Z_0(L)} \; \prod_{p\in \mathscr{E}} \hat{a}_{p,\uparrow}^\dagger|0\cdots 0 \rangle, \label{psi00}
\end{equation}
where $\hat{a}_{p,\uparrow}^\dagger={\hat c}_{p,\uparrow}^\dagger-\nu \sum_{u\in \mathscr{I},(|p-u|=1/2)}{\hat c}_{u,\uparrow}^\dagger$ and $|0\cdots 0\rangle$ denotes the fermionic Fock vacuum state. Here, we remark that  the highest weight state $|\psi_0\rangle$ under OBCs takes the same form as (\ref{psi00}), but the definition of $\hat{a}_{p,\uparrow}^\dagger$ changes accordingly. Note that $Z_0(L)$ is the normalization factor to ensure that $|\psi_0\rangle$ is normalized, which depends on what types of the boundary conditions are adopted. Indeed, it is tedious but straightforward to evaluate  $Z_0(L)$ from the explicit expression for $|\psi_0\rangle$. Under PBCs, we have
\begin{widetext} 
	
	\begin{equation*} 
		Z_0(L)=\sqrt{(1+2\nu^2)^{L/2}+ L\sum_{r=1}^{[L/4]} \left(\frac{(-1)^r}{2r}C_{L/2-r-1}^{r-1}\nu^{4r}(1+2\nu^2)^{L/2-2r}\right)+(-1)^{L/2+1}2\nu^{L}}.
	\end{equation*}
\end{widetext}
Here $[L/4]$ is an integer that is the closest to but less than $L/4$.
Under OBCs, we have
\begin{equation*} 
	Z_0(L)=\sqrt{\sum_{r=0}^{L/2}C_{L-r}^r\nu^{2r}}.
\end{equation*}	
An alternative way to derive $Z_0(L)$ from an exact graded MPS representation for  $|\psi_0\rangle$ may be found in Sec.~B of the SM. Indeed, we are led to two distinct but equivalent expressions for $Z_0(L)$ under PBCs.

In particular, Eq.~(\ref{psi00}) for the highest weight state  $|\psi_0\rangle$ at quarter filling may be exploited to check that  the local order parameter $\langle {\hat S}^z_x\rangle$ is non-zero. This is also valid for the highest weight state at a specific filling (away from quarter filling), with its explicit expression presented in Section~\ref{obs}. As demonstrated there,  the values of the filling fraction $f$ form a dense subset in the entire range $[0,1/2]$ in the thermodynamic limit, thus confirming that SSB with one type-B GM occurs at any filling ($f\neq0$).

Note that  the highest weight state    $|\psi_0\rangle$  under OBCs is formally identical to that under PBCs. However, as already mentioned above,  it is necessary to note that both  ${\hat a}_{1,\sigma}$ and  ${\hat b}_{L/2+1/2,\sigma}$ located on the boundary are different, if OBCs are adopted. For later uses, we present their explicit forms: ${\hat a}_{1,\sigma}={\hat c}_{1,\sigma}-\nu{\hat c}_{3/2,\sigma}$ and ${\hat b}_{L/2+1/2,\sigma}={\hat c}_{L/2+1/2,\sigma}+\nu{\hat c}_{L/2,\sigma}$, together with their Hermitian conjugates: ${\hat a}^\dagger_{1,\sigma}=({\hat a}_{1,\sigma})^\dagger$ and ${\hat b}^\dagger_{L/2+1/2,\sigma}=({\hat b}_{L/2+1/2,\sigma})^\dagger$. As it turns out,   the flat-band ferromagnetic Tasaki  model is peculiar, as far as the physics underlyig distinct types of boundary conditions is concerned.

\section{The orthonormal basis states at and away from quarter filling}~\label{obs}
Generically, the orthonormal basis states, which appear to be the simultaneous eigenvectors of the Hamiltonian (\ref{ham}),  the number of electrons ${\hat N}$ and the $z$-component of the total spin ${\hat S}^z$,  may be labeled by the eigenvalues of  ${\hat N}$ and ${\hat S}^z$. 
Actually, the orthonormal basis states at and away from quarter filling are generated from the repeated action of the lowering operator of the symmetry group ${\rm SU}(2)$ on the highest weight state. 
Note that the symmetry group of the ${\rm SU}(2)$ flat-band ferromagnetic Tasaki model is not semi-simple, due to the presence of the ${\rm U}(1)$ symmetry group generated by ${\hat N}$. Hence, there is always a degenerate ground state that acts as the highest weight state in each sector labeled by the number of  electrons. In fact, there is {\it only} one choice for the highest weight state  $|\psi_0\rangle$ at quarter filling.
As we shall show below, it is entangled. As a consequence, the orthonormal basis states thus constructed at quarter filling are not permutation-invariant with respect to the lattice unit cells. In contrast, there are more choices for the highest weight state away from quarter filling, some of which are not necessarily entangled. Hence, 
the orthonormal basis states, permutation-invariant with respect to the emergent larger-size unit cells, exist away from quarter filling. This difference is a peculiar feature that constitutes a marked difference between the flat-band ferromagnetic Tasaki model and all other known quantum many-body systems undergoing SSB with type-B GMs, given that the permutation invariance implies the absence of the contribution from the area law~\cite{arealaw1,arealaw2,arealaw3}  to the entanglement entropy~\cite{2dtypeb}. This in turn may be attributed to the fact that the boundary between the block and the environment is not well-defined, since its presence is not compatible with the permutation invariance.

At quarter filling ($f=1/2$), a sequence of highly degenerate ground states $|L,M\rangle$ are generated from the repeated action of the lowering operator ${\hat S}^-$ on the highest weight state $|\psi_0\rangle$, which has been presented in Eq.(\ref{psi00}). Hence, we have
\begin{equation}
	|L,M\rangle=\frac{1}{Z(L,M)}({\hat S}^-)^M\;|\psi_0\rangle,
\end{equation}
where $M=0,1,\ldots, L/2$, and  $Z(L,M)$  is introduced to ensure that $|L,M\rangle$ is normalized, which takes the form 
\begin{equation*}
	Z(L,M)=M!\sqrt{C_{L/2}^M}.\label{zlm}
\end{equation*}
Note that the normalization factor $Z(L,M)$ is formally identical to that for the spin-$1/2$ ${\rm SU}(2)$ ferromagnetic Heisenberg model, which acts as the effective spin model at quarter filling  when the on-site repulsive Coulomb interaction is sufficiently strong~\cite{penc}. However, as discussed in Sections~\ref{esd} and~\ref{gmps}, the entanglement entropy for the flat-band ferromagnetic Tasaki model at quarter filling behaves quite differently from that for the spin-$1/2$ ${\rm SU}(2)$ ferromagnetic Heisenberg model, because the highest weight state $|\psi_0\rangle$ is entangled.

As a consequence, it is not permutation-invariant under the permutation operations  with respect to the lattice unit cell.  This is in sharp contrast to all other known quantum many-body systems exhibiting SSB with type-B GMs~\cite{FMGM,LLspin1,Golden, finitesize, stsu4, dtmodel}.  However,  away from quarter filling,  one may construct a sequence of  unentangled degenerate ground states with respect to emergent unit cells, each of which  acts as the (periodic) highest weight state $| {\rm \psi_0}\rangle_q$ at a specific filling, with the filling fraction being $f=1/2-1/q$. Here the period  $q$ is equal to the size of the emergent unit cells. In particular, at this filling, we have
\begin{equation}
	| {\rm \psi_0}\rangle_q=\frac{1}{Z_q(L)}\otimes_{l=1}^{L/q}\{\prod_{p=q(l-1)/2+2}^{ql/2} {\hat a}_{p,\uparrow}^\dagger\}_l \;|0\cdots 0\rangle,
\end{equation}
where $q$ is an even integer but not less than 4,  $L$ is a multiple of $q$, and $Z_q(L)$ denotes the normalization factor for the highest weight state $|\psi_0\rangle_q$, which does not depend on what types of boundary conditions are adopted:
\begin{equation*}
Z_q(L)=\left(\sum_{r=0}^{[(q-1)/2]} C_{q-1-r}^r \nu^{2r}\right)^{L/(2q)}.
\end{equation*}
Here,  $[(q-1)/2]$ is an integer that is the closest to but less than $(q-1)/2$, which is identical to $q/2 -1$ for even $q$.
 
A sequence of highly degenerate ground states $|L,M\rangle_q$ are generated from the repeated action of the lowering operator ${\hat S}^-$ on the highest weight state $|\psi_0\rangle_q$:
\begin{equation}
	|L,M\rangle_q=\frac{1}{Z_q(L,M)}({\hat S}^-)^M\;|\psi_0\rangle_q,
\end{equation}
where $M=0,\;1,\;\ldots,\;L/2-L/q$, and  $Z_q(L,M)$  is introduced to ensure that $|L,M\rangle_q$ is normalized, which takes the form 
\begin{equation*}
	Z_q(L,M)=M!\sqrt{C_{L/2-L/q}^M}.\label{zlm}
\end{equation*}
In particular, note that $q=4$ corresponds to $1/8$ filling: $f=1/4$. Note that if one restricts to the emergent unit cells consisting  of $q$ consecutive lattice sites,  then the filling fraction is $f=1/2-1/q$. However, they are not sufficient to yield a dense subset of the values of the filling fraction $f$ to cover the entire range $[0,1/2]$.  Physically, such a dense subset is crucial to ensure that the entire range of the filling fraction $f$ is covered, which in turn is relevant to the existence of an intrinsic abstract fractal~\cite{hqzhou}, as reflected in the presence of  the golden spiral asymptotically for the ground-state degeneracies under PBCs  and OBCs, which will be discussed in Section~\ref{gsd}. 

Mathematically, it is straightforward to extend our construction described above to more complicated types of emergent unit cells. 
The simplest choice is the emergent unit cells consisting of $q_1+1$ consecutive lattice sites, on which 
$q_1/2$ consecutive $\hat{a}_{p}^{\dagger}$ operators act, and $q_2-1$ consecutive lattice sites, which are in the
fermionic Fock vacuum states, thus yielding the (periodic) highest weight state $|\psi_0\rangle_{q_1,q_2}$ at a specific filling, with  $f=q_1/(2(q_1+q_2))$. Here the period  $q_1+q_2$ is equal to the size of the emergent unit cells, with $q_1$ and $q_2$ chosen to be even.
We thus have 

\begin{equation}
	|\psi_0\rangle_{q_1,q_2}\!=\!\frac{1}{Z_{q_1,\!q_2}(L)}\!\otimes_{l=1}^{\frac{L}{q_1+q_2}}\!\{\prod_{p\!=\!(q_1+q_2)(l\!-\!1)/2+q_2/2+1}^{(q_1+q_2)l/2} \!{\hat a}_{p,\uparrow}^\dagger\}_l|0\cdots 0\rangle,
\end{equation}
where $Z_{q_1,q_2}(L)$ denotes the normalization factor for the highest weight state $|\psi_0\rangle_{q_1,q_2}$, which does not depend on what types of boundary conditions are adopted:
\begin{equation*}
	Z_{q_1,q_2}(L)=\left(\sum_{r=0}^{q_1/2} C_{q_1+1-r}^r \nu^{2r}\right)^{L/(2(q_1+q_2))}.
\end{equation*}

A sequence of highly degenerate ground states $|L,M\rangle_{q_1,q_2}$ are generated from the repeated action of the lowering operator ${\hat S}^-$ on the highest weight state $|\psi_0\rangle_{q_1,q_2}$:
\begin{equation}
|L,M\rangle_{q_1,q_2}=\frac{1}{Z_{q_1,q_2}(L,M)}({\hat S}^-)^M\;|\psi_0\rangle_{q_1,q_2},
\end{equation}
where $M=0,1,\ldots, q_1L/(2(q_1+q_2))$, with $L$ being a multiple of $q_1+q_2$, and $Z_{q_1,q_2}(L,M)$ is introduced to ensure that $|L,M\rangle_{q_1,q_2}$ is normalized, which takes the form 
\begin{equation*}
Z_{q_1,q_2}(L,M)=M!\sqrt{C_{q_1L/(2(q_1+q_2))}^M}.\label{zlmq1q2}
\end{equation*}
Here, $|L,M\rangle_{q_1,q_2}$ are permutation-invariant under the permutation operations with respect to the emergent unit cells consisting of $q_1+q_2$ consecutive lattice sites, as a result of the fact that  the highest weight state $|\psi_0\rangle_{q_1,q_2}$ is unentangled for $f=q_1/(2(q_1+q_2))$.

A few remarks are in order. First,  there is an overlap between the two sequences $|L,M\rangle_q$ and $|L,M\rangle_{q_1,q_2}$. Indeed, if the condition $q_2/q_1=4/(2q-4)$ is satisfied, then we have
 $|\psi_0\rangle_{q_1,q_2}\equiv|\psi_0\rangle_{q}$. In particular, $|\psi_0\rangle_{2,2}\equiv|\psi_0\rangle_{4}$ is the simplest example.
Second, quarter filling may be approached as closely as possible, as $q$ tends to infinity for $| {\rm \psi_0}\rangle_q$ or as $q_1$ tends to infinity, with $q_2$ fixed, for $| {\rm \psi_0}\rangle_{q_1,q_2}$. Third, one may construct 
more complicated emergent unit cells, involving integers $q_1,\ldots,q_\mu$, where $\mu$ is an integer, such that the size of the emergent unit cells  is $q_1+\ldots +q_\mu$, as long as the system size $L$ is an integer multiple of  $q_1+\ldots +q_\mu$. However, as it is readily seen from the expression of the filling fraction $f=q_1/(2(q_1+q_2))$, $f$ involves the ratio of two arbitrary integers in the thermodynamic limit. Hence, the values of $f$ form a dense subset in the interval $[0,1/2]$. In this sense, it is sufficient to restrict to these two types of emergent unit cells. Fourth, it is the emergence of generic (periodic) highest weight states at different fillings that are responsible for the exponential ground-state degeneracies under PBCs and OBCs, thus leading to the non-zero residual entropy, in the sense that the residual entropy measures the disorder present in a specific emergent unit cell, as far as the spin configuration in such an emergent unit cell is concerned. In fact, the disorder has already been reflected in the observation that  
there are many different choices for $q_1,\ldots, q_\mu$ if one fixes the size of the emergent unit cells $q_1+\ldots +q_\mu$.
Fifth,  the highest weight state $|\psi_0\rangle$ at quarter filling is entangled, but  $|\psi_0\rangle_q$ or  $|\psi_0\rangle_{q_1,q_2}$ away from quarter filling are always unentangled for any values of $q$ or $q_1$ and $q_2$, we are led to conclude that there is a marked difference between quarter filling and any other filling. This difference is also reflected in distinct behaviors of the entanglement entropy at and away from quarter filling, as will be discussed in Section~\ref{esd} and Section\ref{ee}. Hence, we encounter a singular limit at quarter filling.  Physically, this is relevant to the fact that, {\it only} at quarter filling, the  flat-band ferromagnetic Tasaki model exhibits the saturated ferromagnetism~\cite{tasaki1,tasakibook}.

\section{The ground-state degeneracies under periodic and open boundary conditions}~\label{gsd}

Given that the Hamiltonian (\ref{ham}) commutes with ${\hat N}$, the Hilbert space is decomposed into the sectors labeled by the number of electrons $N$. Hence, it is convenient to introduce the ground-state degeneracy  under PBCs, denoted as ${\rm dim }(\Omega_L^{\rm PBC})(N)$,  in the sector with $N$ fixed.  An interesting observation is that the Hamiltonian (\ref{ham}) is pathological, in the sense that the lowest energy eigenvectors in the sectors with different $N$ are degenerate. It thus makes sense to define the total ground-state degeneracy ${\rm Dim }(\Omega_L^{\rm PBC})$  under PBCs as the sum over all the sectors with different eigenvalues of ${\hat N}$ that contain degenerate ground states.  That is, we have ${\rm Dim }(\Omega_L^{\rm PBC})=\sum_{N=0}^{L/2}{\rm dim }(\Omega_L^{\rm PBC})(N)$, given that degenerate ground states are absent in the sectors when $N$ is beyond $L/2$, since double occupation at one lattice site is energetically unfavorable due to the on-site repulsive interaction. 

At quarter filling ($f=1/2$), we have ${\rm dim }(\Omega_L^{\rm PBC})(L/2)=L/2+1$~\cite{tasaki1};  
away from quarter filling ($0\le N<L/2$), we have 
${\rm dim }(\Omega_L^{\rm PBC})(N)=L/(L-N)C_{L-N}^N$~\cite{tasakidegeneracy}, where $C_{L-N}^{N}=(L-N)!/(N!(L-2N)!)$ is the binomial coefficient. Hence, for $L\ge 4$, the ground-state degeneracy ${\rm Dim }(\Omega_L^{\rm PBC})$ under PBCs takes the form
\begin{equation} 
	{\rm Dim }(\Omega_L^{\rm PBC})=(\frac{3+\sqrt{5}}{2})^{L/2} + (\frac{3-\sqrt{5}}{2})^{L/2} +L/2-1. \label{rrpbc}
\end{equation}

\begin{table}
	\renewcommand{\arraystretch}{1.5}
	\begin{tabular}{c|cccccc}
		\hline \hline
		$L$\;\;\; \;&\;\;\; 4\;\;\;\;\;& \;\;\; 6\;\;\;\;&\;\;\;\; 8 \;\;\;\;&\;\;\;10\;\;\;\;& \;\;\;\;12 \\
		\hline 
		$E_{gs}$\;\;\;&\;\;\;0\;\;\;\;&\;\;\;\; 0 \;\;\;\;&\;\;\;\;0\;\;\;\;&\;\;\;\;0\;\;\;\;&\;\;\;\;0   \\
		${\rm Dim }(\Omega_L^{\rm PBC})$\;\;\;&\;\;\;8\;\;\;\;&\;\;\;\;20\;\;\;\;&\;\;\;\;50\;\;\;\;&\;\;\;\;127\;\;\;\;&\;\;\;\;327\\
		${\rm Dim }(\Omega_L^{\rm OBC})$\;\;\;&\;\;\;8\;\;\;\;&\;\;\;\;21\;\;\;\;&\;\;\;\;55\;\;\;\;&\;\;\;\;144\;\;\;\;&\;\;\;\;377\\
		\hline \hline
	\end{tabular}
	\caption{The ground-state energy $E_{gs}$ and the ground-state degeneracies, ${\rm Dim }(\Omega_L^{\rm PBC})$ and ${\rm Dim }(\Omega_L^{\rm OBC})$, from the exact diagonalization for the one-dimensional flat-band ferromagnetic Tasaki model under PBCs and OBCs,  where the system size $L$ is up to $L=12$. Here, typical values of the parameters in the Hamiltonian (\ref{ham}) have been chosen as $t=1$, $U=1$ and  $\nu=0.1$. However, our numerical examination confirms that the ground-state degeneracies ${\rm Dim }(\Omega_L^{\rm PBC})$ and ${\rm Dim }(\Omega_L^{\rm OBC})$ do not depend on any specific choice of the parameters.}
	\label{tab:gs}  
\end{table}
A derivation of Eq.(\ref{rrpbc}) has been sketched in Sec.~A of the Supplementary Material (SM).
A similar definition is also valid under OBCs. As argued in  Sec.~A of the SM,  it is found that for $L\ge 4$, the ground-state degeneracy  under OBCs  takes the form
\begin{align}
	{\rm Dim }(\Omega_L^{\rm OBC}) = \frac{(3+\sqrt 5)^{L/2} - (3-\sqrt 5)^{L/2}}{ 2^{L/2} \sqrt{5}  }.\label{rrobc}
\end{align}

An intriguing observation is that both  ${\rm Dim }(\Omega_L^{\rm PBC})$ and ${\rm Dim }(\Omega_L^{\rm OBC})$ constitute (a modified form of) the Fibonacci sequences. In particular, the golden ratio $\phi  = (\sqrt{5}+1)/2$ appears in both (\ref{rrpbc})  and (\ref{rrobc}).
In fact, both (\ref{rrpbc})  and (\ref{rrobc})  are exponential with the system size $L$. Indeed, if the system size $L$ is large enough, then both  of them behave asymptotically as the golden spiral - a self-similar geometric object. Hence, the residual entropy per the lattice unit cell is non-zero in the grand canonical ensemble, when the thermodynamic limit is approached. In fact, the appearance of the golden ratio $\phi$ in  the non-zero residual entropy for this model under PBCs has already been found in Ref.~\cite{goldenratio}.

Here, we remark that the above analytical results for the ground-state degeneracies ${\rm Dim }(\Omega_L^{\rm PBC})$ and ${\rm Dim }(\Omega_L^{\rm OBC})$
have been confirmed from the exact diagonalization. The ground-state energy and the ground-state degeneracies have been listed in Table~\ref{tab:gs}  under PBCs and OBCs, where the system size $L$ is up to $L=12$ (also cf. Table~\ref{tab:gsobc} in Sec.~A of the SM).

\section{Exact Schmidt decompositions and self-similarities}~\label{esd}

Although degenerate ground states $|L,M\rangle$ ($M=0,1,\ldots,N$) at quarter filling admit an exact Schmidt decomposition, the self-similarities in the real space are lost, if one is restricted to the sector at this fixed filling. Indeed, if the system is partitioned into a block ${\cal B}$  consisting of $n/2$ lattice unit cells and an environment ${\cal E}$  consisting of the remaining $(L-n)/2$ lattice unit cells, then the block $\cal B$ and the environment $\cal E$, as a subsystem, do not share the same type of the orthonormal basis states as the entire system. Mathematically, this stems from the fact that if the entire system  is at quarter filling, then it is impossible for the subsystems to be restricted to this fixed filling, once the bipartition is performed. This is relevant to the fact that
the highest weight state $|\psi_0\rangle$ admits an exact graded MPS representation, with the bond dimension being two, as discussed in 
Section~\ref{gmps}. More precisely, one may perform an exact Schmidt decomposition for the highest weight state $|\psi_0\rangle$ at quarter filling, which requires that the Schmidt basis states for both the block and the environment are not at quarter filling (for an explicit construction of  the Schmidt basis states for both the block and the environment, cf. Sec. B of the SM).

In contrast,  more degenerate ground states exist that may act as the highest weight state at a specific filling, if the system is away from quarter filling.
However,  {\it only} one of these degenerate ground states  is periodic. Here, we restrict ourselves to a sequence of highly degenerate ground states  $|L,M\rangle_q$ and $|L,M\rangle_{q_1,q_2}$ that are generated from the periodic highest weight state $| {\rm \psi_0}\rangle_q$ and $| {\rm \psi_0}\rangle_{q_1,q_2}$, with the period being the size of emergent unit cells. Here we emphasize that the numbers of emergent unit cells in both the block and the environment are always integers such that $|L,M\rangle_q$ and $|L,M\rangle_{q_1,q_2}$ are not entangled with respect to the emergent unit cells. 
It is found that degenerate ground states $|L,M\rangle_q$ admit an exact Schmidt decomposition
\begin{equation}
	|L,M\rangle_q\!=\! \sum\limits_{k=0}^{\min(n/2-n/q,M)}\lambda_q(L,n,M,k)
	|n,k\rangle_q|L-n,M-k\rangle_q. \label{schmidtq}
\end{equation}
Here, $\lambda_q(L,n,M,k)$ denote the Schmidt coefficients, taking the form
\begin{equation*}
	\lambda_q(L,n,M,k)=C_{M}^{k}\frac{Z_q(n,k)Z_q(L-n,M-k)}{Z_q(L,M)},
	\label{lambdanmk}
\end{equation*}
where $Z_q(n,k)$, $Z_q(L-n,M-k)$ and $Z_q(L,M)$ are normalization factors of $|n,k\rangle_q$, $|L-n,M-k\rangle_q$ and $|L,M\rangle_q$, respectively. 
This decomposition reflects the self-similarities in the real space, thus implying an abstract fractal underlying the ground-state subspace~\cite{hqzhou}. This is due to the fact that the block $\cal B$ and the environment $\cal E$, as a subsystem, share the same type of orthonormal basis states as that for the entire system at exactly the same filling.  

Similarly, degenerate ground states $|L,M\rangle_{q_1,q_2}$ admit an exact Schmidt decomposition
\begin{align}
|L,M\rangle_{q_1,q_2}= &\sum\limits_{k=0}^{\min(q_1n/(2(q_1+q_2)),M)}\lambda_{q_1,q_2}(L,n,M,k)\times\nonumber \\
&
|n,k\rangle_{q_1,q_2}|L-n,M-k\rangle_{q_1,q_2}. \label{schmidt}
\end{align}
Here, $\lambda_{q_1,q_2}(L,n,M,k)$ denote the Schmidt coefficients, taking the form
\begin{equation*}
\lambda_{q_1,q_2}(L,n,M,k)=C_{M}^{k}\frac{Z_{q_1,q_2}(n,k)Z_{q_1,q_2}(L-n,M-k)}{Z_{q_1,q_2}(L,M)},
\label{lambdanmkq1q2}
\end{equation*}
where $Z_{q_1,q_2}(n,k)$, $Z_{q_1,q_2}(L-n,M-k)$ and $Z_{q_1,q_2}(L,M)$ are normalization factors of $|n,k\rangle_{q_1,q_2}$, $|L-n,M-k\rangle_{q_1,q_2}$ and $|L,M\rangle_{q_1,q_2}$, respectively. 
Recall that
the reduced density matrix $\rho_Q (L,n,M)$ is determined from tracing out the degrees of freedom in the environment ${\cal E}$, where $Q=q$ and $q_1,q_2$. Hence, the entanglement entropy  $S_Q(L,n,M)=-\sum_\kappa\Lambda_Q(L,n,M)\log_2\Lambda_Q(L,n,M)$, is determined from the eigenvalues $\Lambda_Q(L,n, M)$ of the reduced density matrix $\rho_Q (L,n,M)$ with $\Lambda_Q(L,n,M)=[\lambda_Q(L,n,M)]^2$.

Note that  the Schmidt coefficients $\lambda_q(L,n,M,k)$ and $\lambda_{q_1,q_2}(L,n,M,k)$ do not depend on the parameters $t$, $U$ and  $\nu$.
We stress that a singularity already manifests itself in the differences between the exact Schmidt decompositions for the orthonormal basis states at and away from quarter filling, as will be discussed in Section~\ref{ee}. Mathematically, this results from the fact that for $|L,M\rangle_Q$, with $Q \equiv q$ for  $S_q(L,n,M)$ or $Q \equiv q_1,q_2$ for  $S_{q_1,q_2}(L,n,M)$, the orthonormal basis states $|n,k\rangle_Q$ and $|L-n,M-k\rangle_Q$ for the subsystems share the same filling, in contrast to what happens at the quarter filling. However,  the self-similarities in the real space are  still reflected in the exact Schmidt decompositions for the orthonormal basis states, if all values of the filling fraction $f$ are taken into account. The simplest case is exemplified in the Schmidt decomposition for the highest weight state at quarter filling (cf. Sec. B of the SM). In this regard, the presence of other non-periodic degenerate ground states that act as highest weight states is crucial. In other words, an abstract fractal underlying the ground-state subspace exists if all the sectors with different eigenvalues of ${\hat N}$ are taken into account, as already discussed for quantum many-body spin systems undergoing SSB with type-B GMs~\cite{hqzhou}. This has been partially reflected in the observation that the ground-state degeneracies under both PBCs and OBCs behave asymptotically as the golden spiral - a well-known self-similar geometric object.

\section{An exact graded MPS representation for highly degenerate ground states}~\label{gmps}

The presence of the exact Schmidt decompositions for the orthonormal basis states arising from SSB with type-B GMs implies that they admit an exact graded MPS representation, which is adapted from the standard MPS representation for quantum many-body spin systems (for graded tensor network representations, we refer to Refs.~\cite{gradpeps,kraus,fermionpeps,fermionpeps1}; also cf. Sec.\;B  of the SM). 
Here, we restrict ourselves to the orthonormal basis states $|L,M\rangle$ at quarter filling, $|L,M\rangle_4$ at $1/8$ filling and $|L,M\rangle_{2,4}$ at $1/12$ filling. The detailed construction of the graded MPS representations under both PBCs and OBCs has been presented in Sec.\;B  of the SM. 

The procedure consists of five steps as follows. First,  represent the fermionic Fock vacuum state $|0\cdots0\rangle$ in a trivial graded MPS representation, where the bond dimension is one, with its parity being even. Second,
$\hat{a}_{p,\uparrow}^\dagger$  is represented in terms of a graded matrix product operator (MPO), where the bond dimension is two, with one of the basis states in the auxiliary space being even and the other odd. Third,  contract a sequence of the graded MPO  representations for $\hat{a}_{p,\uparrow}^\dagger$'s with a trivial graded MPS representation for the fermionic Fock vacuum state $|0\cdots0\rangle$, thus yielding an exact graded MPS representation for $|\psi_0\rangle$ at quarter filling, for $|\psi_0\rangle_4$ at $1/8$ filling and for $|\psi_0\rangle_{2,4}$ at $1/12$ filling.  Fourth, a power of the lowering operator $({\hat S}^-)^M$  is expressed in terms of a graded MPO representation, a variant of the MPO representation for $({\hat S}^-)^M$ in the ${\rm SU}(2)$ ferromagnetic Heisenberg model~\cite{exactmps}, which in turn is adapted from a MPO representation for a generic operator~\cite{yuping}. Fifth, contract the graded MPO representation for $({\hat S}^-)^M$  with the graded MPS representation for the highest weight state, thus yielding an exact graded MPS representation for degenerate ground states at a specific filling. We shall choose $|\psi_0\rangle$, $|\psi_0\rangle_4$ and $|\psi_0\rangle_{2,4}$ to yield an graded MPS representation for $|L,M\rangle$ at quarter filling,  $|L,M\rangle_4$ at $1/8$ filling and $|L,M\rangle_{2,4}$ at $1/12$ filling.
 
At quarter filling,  an exact graded MPS representation for $|L,M\rangle$ under PBCs takes the form
\begin{widetext}
	\begin{equation}
		|L,M\rangle = \sum_{\alpha_1,\alpha_{3/2},\ldots,\alpha_{p},\alpha_{p+1/2},\ldots, \alpha_{L/2},\alpha_{L/2+1/2}}  C(X^{\alpha_1}F^{\alpha_{3/2}}\cdots E^{\alpha_{p}}F^{\alpha_{p+1/2}}\cdots E^{\alpha_{L/2}}Y^{\alpha_{L/2+1/2}})|\alpha_1\alpha_{3/2}\cdots\alpha_{p}\alpha_{p+1/2}\cdots\alpha_{L/2}\alpha_{L/2+1/2}\rangle,
		\label{lmDEFG}
	\end{equation}
\end{widetext}
where  $X^{\alpha_1}$ and $Y^{\alpha_{L/2+1/2}}$ are the two tensors at the two ends,
and  $E^{\alpha_{p}}$ ($p=2,\;3,\;\ldots, \;L/2$) and  $F^{\alpha_{u}}$ ($u=3/2,\;5/2,\; \ldots, \;L/2-1/2$) are the tensors in the bulk. Their  explicit expressions may be found in Sec.~B of the SM.
Here, $\alpha_p$ ($p\in \mathscr{E}$) and $\alpha_u$ ($u\in\mathscr{I}$) are physical indices representing the four basis states $|0\rangle_x$, $|\uparrow\rangle_x$, $|\downarrow\rangle_x$ and $|\uparrow\downarrow\rangle_x$ at each lattice site $x$, with $|0\rangle_x$ and $|\uparrow\downarrow\rangle_x$ being even and $|\uparrow\rangle_x$, $|\downarrow\rangle_x$ being odd,  respectively.  In addition,  the symbol $C$ has been exploited to represent a contraction operation, formally identical to the  supertrace or trace in the auxiliary graded vector spaces, up to an overall minus sign, which depends on the values of the system size $L$.  Note that  we only need to keep this contraction operation $C$ when the bond dimension is greater than one under PBCs. Indeed, as seen below for $1/8$ filling and $1/12$ filling, the bond dimension is one for the bonds between the emergent unit cells, so the contraction operation $C$ has been dropped off under PBCs. 

At quarter filling,  an exact graded MPS representation for $|L,M\rangle$ under OBCs takes the form
\begin{widetext}
	\begin{equation}
	|L,M\rangle = \sum_{\{\alpha_x\},\;x\in \Lambda}  (U^{\alpha_1}F^{\alpha_{3/2}}\cdots E^{\alpha_{p}}F^{\alpha_{p+1/2}}\cdots E^{\alpha_{L/2}}V^{\alpha_{L/2+1/2}})|\alpha_1\alpha_{3/2}\cdots\alpha_{p}\alpha_{p+1/2}\cdots\alpha_{L/2}\alpha_{L/2+1/2}\rangle,
	\label{lmDEFGobc}
	\end{equation}
\end{widetext}
where the two tensors $U^{\alpha_1}$ and $V^{\alpha_{L/2+1/2}}$ are the two tensors at the two ends different from $X^{\alpha_1}$ and $Y^{\alpha_{L/2+1/2}}$ in Eq.~(\ref{lmDEFG}). Their explicit expressions may be found in Sec.~B of the SM.

At $1/8$ filling, an exact graded MPS representation for $|L,M\rangle_4$  takes the form
\begin{widetext}
	\begin{equation}
		|L,M\rangle_4 = \sum_{\alpha_1,\ldots,  \alpha_{p}, \alpha_{p+1/2}, \atop
			 \alpha_{p+1}, \alpha_{p+3/2},\ldots, \alpha_{L/2+1/2}}  (X^{\alpha_{1}}\cdots E^{\alpha_{p}}F^{\alpha_{p+1/2}}G^{\alpha_{p+1}}H^{\alpha_{p+3/2}}\cdots Y^{\alpha_{L/2+1/2}})|\alpha_1\cdots\alpha_{p} \alpha_{p+1/2}\alpha_{p+1} \alpha_{p+3/2}\cdots \alpha_{L/2+1/2}\rangle,
	\label{lm4DEFGH}
	\end{equation}
\end{widetext}
where  $X^{\alpha_1}$ and $Y^{\alpha_{L/2+1/2}}$ are the two tensors at the two ends,
and  $E^{\alpha_{p}}$ ($p=3,\;5,\;\ldots, \;L/2-1$),  $F^{\alpha_{u}}$ ($u=3/2,\;7/2,\;\ldots, \;L/2-1/2$),  $G^{\alpha_{p}}$($p=2,\;4,\;\ldots, \;L/2$) and $H^{\alpha_{u}}$ ($u=5/2,\;9/2,\;\ldots, \;L/2-3/2$) are the tensors in the bulk. Their  explicit expressions may be found in Sec.~B of the SM.  
Note that $|L,M\rangle_4$ are permutation-invariant with respect to the emergent unit cells, which do not depend on what types of boundary conditions are adopted.

At $1/12$ filling, an exact graded MPS representation for $|L,M\rangle_{2,4}$  takes the form
\begin{widetext}
	\begin{align}
	|L,M\rangle_{2,4} = \sum_{\alpha_1,\ldots,  \alpha_{p}, \alpha_{p+1/2}, \alpha_{p+1}, \alpha_{p+3/2},\atop \alpha_{p+2}, \alpha_{p+5/2},\ldots, \alpha_{L/2+1/2}}  (X^{\alpha_{1}}\cdots E^{\alpha_{p}}E^{\alpha_{p+1/2}}E^{\alpha_{p+1}}F^{\alpha_{p+3/2}}G^{\alpha_{p+2}}H^{\alpha_{p+5/2}}\cdots Y^{\alpha_{L/2+1/2}})\nonumber \\
	|\alpha_1\cdots\alpha_{p} \alpha_{p+1/2}\alpha_{p+1} \alpha_{p+3/2}\alpha_{p+2} \alpha_{p+5/2}\cdots \alpha_{L/2+1/2}\rangle,
	\label{lm24DEFGH}
	\end{align}
\end{widetext}
where $X^{\alpha_1}$ and $Y^{\alpha_{L/2+1/2}}$ are the two tensors at the two ends,
and $E^{\alpha_{p/u}}$ ($p=2,\;4,\;5,\;\ldots, \;L/2-2$ and $u=3/2,\;9/2,\;\ldots, \;L/2-3/2$),  $F^{\alpha_{u}}$ ($u=5/2,\;11/2,\;\ldots, \;L/2-1/2$),  $G^{\alpha_{p}}$($p=3,\;6,\;\ldots, \;L/2$) and $H^{\alpha_{u}}$ ($u=7/2,\;13/2,\;\ldots, \;L/2-5/2$) are the tensors in the bulk. Their  explicit expressions may be found in Sec.~B of the SM.  
Note that $|L,M\rangle_{2,4}$ are permutation-invariant with respect to the emergent unit cells, which do not depend on what types of boundary conditions are adopted.

We remark that an exact graded MPS representation for the orthonormal basis states provides an alternative powerful means to investigate the emergence of ferromagnetism, in particular  when a perturbation is introduced into the flat-band ferromagnetic Tasaki model, with the frustration-free nature of the Hamiltonian being left intact under perturbation so that the nearly-flat band becomes dispersive~\cite{nearlyflat,nearlyflat1}.

\section{A finite system-size scaling analysis of the entanglement entropy at and away from quarter filling}~~\label{ee}

We turn to a finite system-size scaling analysis of the entanglement entropy for the flat-band ferromagnetic Tasaki model, when it is at and away  from quarter filling. In our numerical implementation,  the values of the parameters in the Hamiltonian (\ref{ham}) are chosen as $t=1$, $U=1$ and  $\nu=0.1$, which may be regarded as their typical values. However, we stress that the entanglement entropy $S_Q(L,n,M)$  for the orthonormal basis states $|L,M\rangle_Q$ do not depend on  the values of the parameters, as follows from the independence of the Schmidt coefficients $\lambda_{Q}(L,n,M,k)$ ($Q=q$ and $q_1,q_2$) on the parameters, in contrast to the entanglement entropy $S(L,n,M)$  for the orthonormal basis states $|L,M\rangle$.

Away from quarter filling, if the system size $L$ is finite, then the entanglement entropy $S_Q(L,n,M)$ for the orthonormal basis state $|L,M\rangle_Q$ ($M \neq 0$) scales  with the block size $n$ as follows~\cite{finitesize}, 
\begin{equation}
 S_Q(L,n,M)=\frac {N_B}{2}\log_2 \left( \frac{\frac{n}{2}(\frac{L}{2}-\frac{n}{2})}{\frac{L}{2}}\right)+S_{Q0}(L,M),~\label{srq}
\end{equation}
where $S_{Q0}(L,M)$ denotes a non-universal additive constant and the number of type-B GMs $N_B$ is one for the flat-band ferromagnetic Tasaki model (\ref{ham}). Here, we emphasize that, as argued in Ref.~\cite{2dtypeb}, the contribution from the area law~\cite{arealaw1,arealaw2,arealaw3} is absent, due to the permutation invariance of the orthonormal basis states $|L,M\rangle_Q$ with respect to the emergent unit cells. 

However, at quarter filling, an extension of the above argument is needed for the orthonormal basis states $|L,M\rangle$. As it turns out, the scaling relation (\ref{srq}) still works for $|L,M\rangle$,
with $S_Q(L,n,M)$ and  $S_{Q0}(L,M)$ being  replaced by $S(L,n,M)$ and $S_{0}(L,M)$, respectively. This is because the area law contribution  from high energy degrees of freedom in the vicinity of the boundary is a constant in one spatial dimension, so it mixes up with other non-universal contributions as an additive constant. However,  $S_{0}(L,M)$ depends on what types of boundary conditions are adopted at quarter filling. The  scaling behaviors of the entanglement entropy for the flat-band ferromagnetic Tasaki model therefore are formally identical to that for quantum many-body spin systems undergoing SSB with type-B GMs.

Note that, in the scaling relation (\ref{srq}), both the block size $n$ and the system size $L$ have been counted in terms of the number of the lattice unit cells contained. In the thermodynamic limit $L \rightarrow \infty$, the entanglement entropy scales with the block size $n$ logarithmically, with the prefactor being half the number of type-B GMs:
\begin{equation}
	S_{Qf}(n)=\frac {N_B}{2}\log_2 \frac{n}{2}+S_{Qf0},~\label{srqn}
\end{equation}
where $S_{Qf}(n)$ and $S_{Qf0}$ denote the limits of $S_Q(L,n,M)$ and $S_{Q0}(L,M)$ as the thermodynamic limit is approached, with fixed non-zero $M/L$. 
Here we stress that the entanglement entropy $S_Q(L,n,M)$ is simply saturated to be a non-zero constant if $M=0$,  as already found in Ref.~\cite{masterthesis} for the highest weight state $|\psi_0\rangle$ at quarter filling. Indeed, the tight upper bound is $2$ under PBCs and $1$ under OBCs, since the bond dimension is two. In fact, this non-zero saturated entanglement entropy for  the highest weight state $|\psi_0\rangle$ at quarter filling is a characteristic feature for the flat-band ferromagnetic Tasaki model, given that the highest weight state is always unentangled for all other known examples of quantum many-body systems exhibiting SSB with type-B GMs~\cite{FMGM,LLspin1,Golden, finitesize, stsu4, dtmodel}.

\begin{figure}[ht!]
	\centering
	\includegraphics[width=0.42\textwidth]{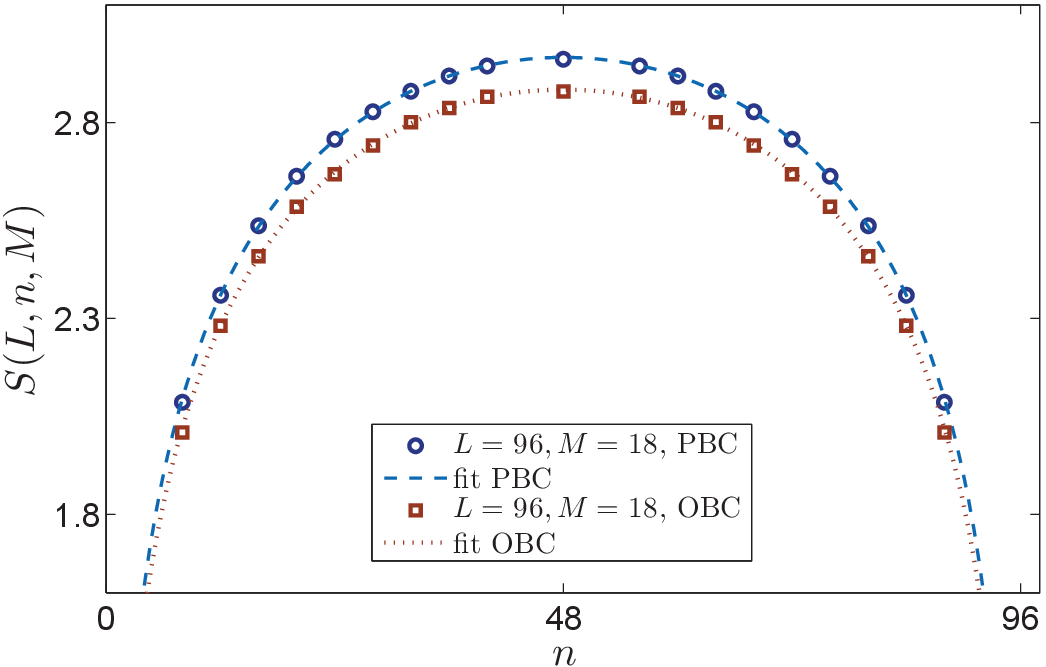}
	\caption{ The entanglement entropy $S(L,n,M)$ versus $n$ for the orthonormal basis states $|L,M\rangle$  in the flat-band ferromagnetic Tasaki model  at quarter filling, when $L=96$ and $M=18$.  Here both PBCs and OBCs  are considered. 
	Our numerical results show that under both PBCs and OBCs, the prefactor is close to the exact value $N_B/2$, where $N_B=1$, with a relative error being less than 2$\%$. In contrast, the non-universal constant  $S_{0}(L,M)$ depends on what types of boundary conditions adopted at quarter filling, thus confirming the contribution from the area law to the entanglement entropy. Here, $n$ is a multiple of two and ranges from $8$ to $88$.
	}
	\label{halffilling}
\end{figure}

In Fig.~\ref{halffilling}, we plot the entanglement entropy $S(L,n,M)$ versus $n$  for the orthonormal basis states $|L,M\rangle$  under both PBCs and OBCs, respectively, at quarter filling. Here we have taken $L=96$ and $M=18$. The prefactor  is close to the exact value $1/2$, with a relative error being less than 2$\%$. In contrast,  $S_{0}(L,M)$ depends on what types of boundary conditions are adopted.
As a result, the entanglement entropy accommodates a contribution from the area law at quarter filling.

\begin{figure}
	\centering
	\includegraphics[width=0.42\textwidth]{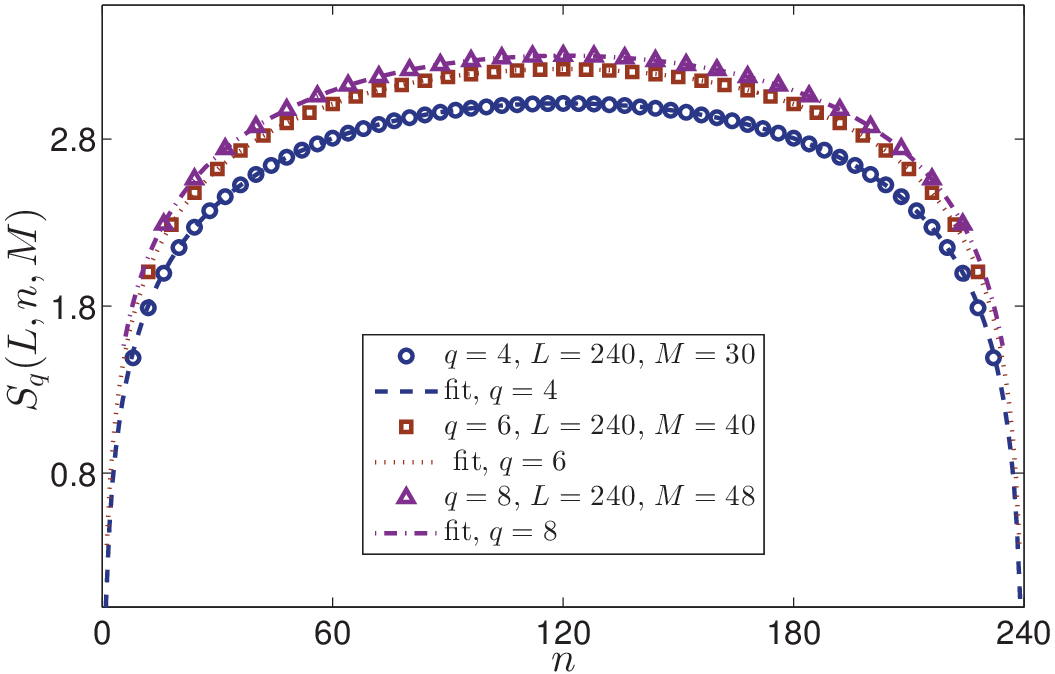}
	\caption{The entanglement entropy $S_q(L,n,M)$ versus $n$ for the orthonormal basis states $|L,M\rangle_q$ in the flat-band ferromagnetic Tasaki model for $q=4,6$ and 8, when $L=240$ and $M=30$, $40$ and $48$, respectively, with the filling fraction $f$ being $1/4$, $1/3$ and $3/8$. The prefactor is close to the exact value $N_B/2$, where $N_B=1$, with a relative error being less than 1$\%$. Here, $n$  is a multiple of $q$ and ranges from $8$ to $232$. 
	}\label{sq}
\end{figure}

In Fig.~\ref{sq}, we plot the entanglement entropy $S_q(L,n,M)$ versus $n$ for the orthonormal basis states $|L,M\rangle_q$  for $q=4,6$ and 8. Here we have taken $L=240$ and $M=30$, $40$ and $48$, respectively. Accordingly, the filling fraction $f$ is $1/4$, $1/3$ and $3/8$, respectively. The prefactor is close to the exact value $1/2$, with a relative error being less than 1$\%$. Here, the non-universal constant  $S_{q0}(L,n,M)$ does not depend on the type of boundary condition are adopted.

	\begin{figure}
	\centering
	\includegraphics[width=0.42\textwidth]{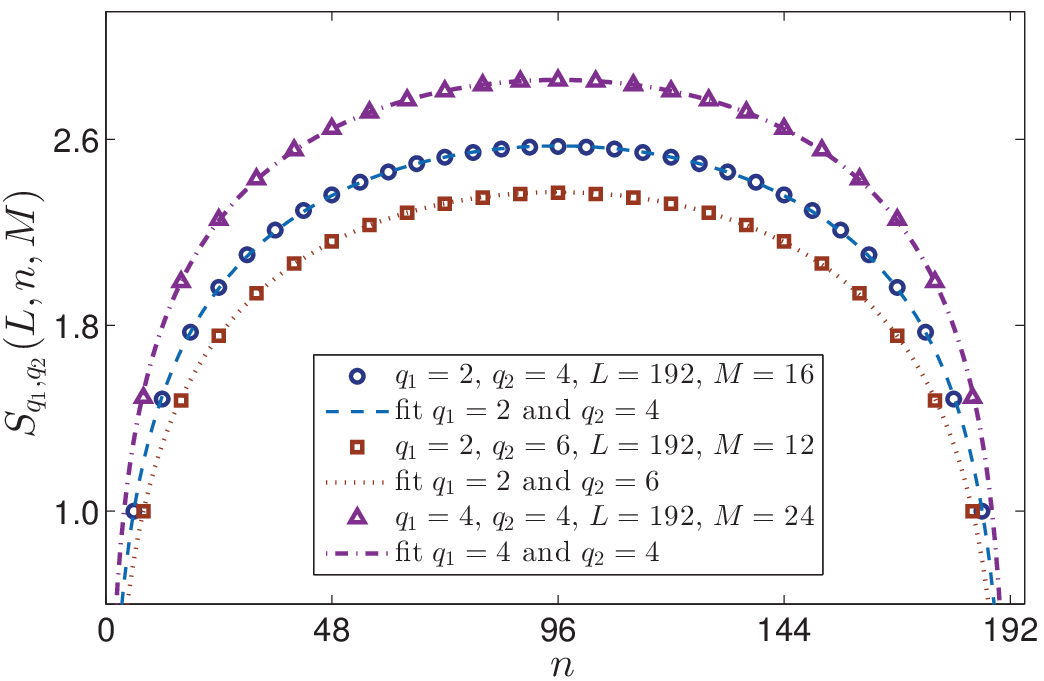}
	\caption{The entanglement entropy $S_{q_1,q_2}(L,n,M)$ versus $n$ for $|L,M\rangle_{q_1,q_2}$ in the one-dimensional Tasaki model for $q_1=2$ and $q_2=4$, $q_1=2$ and $q_2=6$, and $q_1=4$ and $q_2=4$, when $L=192$ and $M=16$, $12$ and $24$, respectively, with the filling fraction $f$ being 1/6, 1/8 and 1/4.  The prefactor is close to the exact value $N_B$/2, where $N_B =1$, with a relative error being less than 3$\%$, 4$\%$ and 2$\%$, respectively, and  the non-universal constant  $S_{q_1,q_2,0}(L,n,M)$ does not depend on what types of boundary conditions adopted.  Here, $n$ is a multiple of $q_1+q_2$ and ranges from $q_1+q_2$ to $L-q_1-q_2$. 
	}\label{sq1q2}
\end{figure}

In Fig.~\ref{sq1q2}, we plot the entanglement entropy $S_{q_1,q_2}(L,n,M)$  versus $n$ for highly degenerate ground states $|L,M\rangle_{q_1,q_2}$ in the one-dimensional Tasaki model for $q_1=2$ and $q_2=4$, $q_1=2$ and $q_2=6$, and $q_1=4$ and $q_2=4$. Here we have taken $L=192$ and $M=16$, $12$ and $24$, respectively.  Accordingly,  the filling fraction $f$ is 1/6, 1/8 and 1/4. The prefactor is close to the exact value $N_B$/2, where $N_B =1$, with a relative error being less than 3$\%$, 4$\%$ and 2$\%$, respectively.  Here, the non-universal constant  $S_{q_1,q_2,0}(L,n,M)$ does not depend on what types of boundary conditions adopted.

We remark that quarter filling is approached, as $q$ tends to infinity or as $q_1$ tends to infinity, with $q_2$ fixed. Hence, the entanglement entropy is singular at quarter filling. Indeed,  $S(L,n,M)$ is not identical to  $S_q(L,n,M)$ as  $q$ tends  to infinity  or to  $S_{q_1,q_2}(L,n,M)$ as $q_1$ tends to infinity, with $q_2$ fixed. 
Hence, the flat-band ferromagnetic Tasaki model is {\it special}, in the sense that the entanglement entropy is subject to the area law, as reflected in the observation that the entanglement entropy for the flat-band ferromagnetic Tasaki model depends on the types of boundary conditions adopted at quarter filling,  but does not necessarily away from quarter filling.

\section{Concluding remarks}

We have carried out a systematic investigation into the one-dimensional  flat-band ferromagnetic Tasaki model from an entanglement perspective. As it turns out,  highly degenerate ground states arise from the SSB pattern: ${\rm SU}(2) \rightarrow {\rm U}(1)$ SSB  with one type-B GM for this strongly correlated itinerant electron model. In fact, the ground-state degeneracies under PBCs and OBCs constitute essentially the Fibonacci sequences, behaving asymptotically as the golden spiral - a self-similar geometric object.
In particular, a finite system-size scaling analysis of the entanglement entropy has been performed for a set of the orthonormal basis states  at and away from quarter filling,  confirming that the entanglement entropy scales logarithmically with the block size in the thermodynamic limit, with the prefactor being half the number of type-B GMs for the orthonormal basis states - highly degenerate ground states generated from the repeated action of the lowering operator of the symmetry group ${\rm SU}(2)$ on the highest weight state at a specific filling. In addition, an exact graded MPS representation has been constructed for the orthonormal basis states, which might provide an alternative powerful means to account for the emergence of ferromagnetism, in particular when a perturbation is introduced into the model so that the nearly-flat band becomes dispersive~\cite{nearlyflat,nearlyflat1}. 

The orthonormal basis states exhibit self-similarities in real space that manifest themselves in an exact Schmidt decomposition  away from quarter filling, with $f=1/2-1/q$.  They are unentangled and permutation-invariant under the permutation operations with respect to an emergent unit cell consisting of $q$ consecutive lattice sites.
This may be extended to an emergent unit cell that involves at least two integers $q_1$ and $q_2$, thus forming a dense subset of the values of the filling fraction $f$ to cover the entire range $[0,1/2]$  in the thermodynamic limit.  Mathematically, the subset of  the values of the filling fraction $f$ is countably infinite, but dense in the entire range, so the subset of the orthonormal basis states constructed at this subset of the values of the filling fraction $f$ is also countably infinite, in contrast to an uncountably infinite set consisting of all the other degenerate ground states, in the thermodynamic limit. Physically, the presence of such a dense subset ensures the existence of an intrinsic abstract fractal~\cite{hqzhou}, as already reflected in the golden spiral describing asymptotically the ground-state degeneracies under PBCs  and OBCs.
In contrast, this feature is lost at quarter filling, since the highest weight state is entangled at quarter filling. As a result, an exact Schmidt decomposition  at quarter filling does not exhibit self-similarities in real space. This is in sharp contrast to other fillings less than quarter. Hence, the singularity has been revealed that implies the presence of a quantum phase transition point at quarter filling, if the filling fraction $f$ is chosen to be a control parameter. This is consistent with the emergence of the saturated ferromagnetism {\it only} at quarter filling for the ${\rm SU}(2)$ flat-band ferromagnetic Tasaki model.

The dependence of the entanglement entropy on the types of boundary conditions adopted at quarter filling originates from the fact that the highest weight state is entangled, so the orthonormal basis states generated are not permutation-invariant  with respect to the lattice unit cell. As a consequence, the contribution from the area law to the entanglement entropy is present~\cite{hqzhou}. Since we are dealing with the one-dimensional version, this contribution appears to be {\it only} a constant, so it mixes up with the non-universal additive constant. One may anticipate that the contribution from the area law does not mix up with the non-universal additive constant in the $d$-dimensional version, which must manifest itself in an exact graded PEPS  representation for the highest weight state at  $1/(2d+2)$ filling, given that a graded PEPS representation obeys the area law, if $d$ is not less than two.

As a closing remark, we mention that the flat-band ferromagnetic Tasaki model may be extended from the  symmetry group ${\rm SU}(2)$ to the symmetry group ${\rm SU}(2s+1)$~\cite{wzhang,katsura,masterthesis}, which is the variant of the  ${\rm SU}(2s+1)$ Hubbard model on the decorated lattices. We expect that SSB with $2s$ type-B GMs emerges in this variant.  The approach developed here is thus amenable to a systematic finite system-size scaling analysis of the entanglement entropy for the orthonormal basis states in the  ${\rm SU}(2s+1)$ flat-band ferromagnetic Tasaki model.

{\it Acknowledgment.-}  Helpful discussions with Murray Batchelor, Hosho Katsura and Jesse Osborne are greatly appreciated. We also thank Hosho Katsura for bringing our attention to Refs.~\cite{katsura,masterthesis}.  I. P. M. acknowledges funding from the National Science and Technology Council (NSTC) Grant No. 122-2811-M-007-044.
When the manuscript was nearly completed, we learned that Kensuke Tamura had investigated the entanglement entropy for the highest weight state 
at quarter filling and found that it is saturated, consistent with our conclusion.

%%%%%%%%%%%%%%%%%%%%%%%%%%%%%%%%%%%%%%%%%%%%%%%%%Appendix%%%%%%%%%%%%%%%%%%%%%%%5
\newpage
\onecolumngrid
\newpage
\section*{Supplementary Material}
\twocolumngrid
\setcounter{page}{1}
\setcounter{equation}{0}
\setcounter{figure}{0}
\setcounter{table}{0}
\renewcommand{\theequation}{S\arabic{equation}}
\renewcommand{\thefigure}{S\arabic{figure}}
\renewcommand{\thetable}{S\arabic{table}}
\renewcommand{\bibnumfmt}[1]{[S#1]}
\renewcommand{\citenumfont}[1]{S#1}

\subsection{A derivation of the ground-state degeneracies under PBCs and OBCs }

Here we sketch a derivation of the ground-state degeneracies under PBCs and OBCs for the one-dimensional flat-band ferromagnetic Tasaki model.

Let us start from  the ground-state degeneracy under PBCs. According to the definition ${\rm Dim }(\Omega_L^{\rm PBC})=\sum_{N=0}^{L/2}{\rm dim }(\Omega_L^{\rm PBC})(N)$, we may split it into two parts ${\rm Dim }(\Omega_L^{\rm PBC})=\sum_{N=0}^{L/2-1}{\rm dim }(\Omega_L^{\rm PBC})(N) + {\rm dim }(\Omega_L^{\rm PBC})(L/2)$. For brevity, we define $G_L = \sum_{N=0}^{L/2-1}{\rm dim }(\Omega_L^{\rm PBC})(N)+2$. We stress that $L$ is always even for the the one-dimensional flat-band ferromagnetic Tasaki model. Given ${\rm dim }(\Omega_L^{\rm PBC})(L/2)=L/2+1$~\cite{smtasaki1} and  
${\rm dim }(\Omega_L^{\rm PBC})(N)=L/(L-N)C_{L-N}^N$~\cite{smtasakidegeneracy} if $0\le N<L/2$, we have ${\rm Dim }\;(\Omega_L^{\rm PBC})=G_L +L/2-1$, when $L$ is even.
For our purpose, it is convenient to introduce $G_L$ when $L$ is odd, which is defined as follows
\begin{align*}
	G_L=	\sum_{N=0}^{(L-1)/2}\frac{L}{L-N}C_{L-N}^N.
\end{align*}
Note that, if $L$ is an even integer, then $G_L$ may be recast into the form
\begin{align*}
	G_L=	\sum_{N=0}^{L/2}\frac{L}{L-N}C_{L-N}^N.%=(\frac{1+\sqrt{5}}{2})^{L} + (\frac{1-\sqrt{5}}{2})^{L}.
\end{align*}
We aim to show that $G_L$ satisfies the three-term recursive relation:	$G_L=G_{L-1}+G_{L-2}$, which is formally identical to that for the Fibonacci sequence. Indeed, for an odd $L$, we have
\begin{widetext}
	\begin{align*}
		G_L=&\sum_{N=0}^{(L-1)/2}\frac{L}{L-N}C_{L-N}^N =1+\sum_{N=1}^{(L-1)/2}\frac{L}{L-N}C_{L-N}^N =1+\sum_{N=1}^{(L-1)/2}\frac{L}{L-N}(C_{L-N-1}^N+C_{L-N-1}^{N-1})\\
     	=&1+\sum_{N=1}^{(L-1)/2}\frac{L}{L-N}C_{L-N-1}^N+\sum_{N=1}^{(L-1)/2}\frac{L}{L-N}C_{L-N-1}^{N-1}=
     	\sum_{N=0}^{(L-1)/2}\frac{L}{L-N}C_{L-N-1}^N+\sum_{N=1}^{(L-1)/2}\frac{L}{L-N}C_{L-N-1}^{N-1} \\
     	=&\sum_{N=0}^{(L-1)/2}(\frac{L}{L-N}-\frac{L-1}{L-N-1})C_{L-N-1}^N+\sum_{N=0}^{(L-1)/2}\frac{L-1}{L-N-1}C_{L-N-1}^N+\\
     	&\sum_{N=1}^{(L-1)/2}(\frac{L}{L-N}-\frac{L-2}{L-2-(N-1)})C_{L-N-1}^{N-1}+\sum_{N=1}^{(L-1)/2}\frac{L-2}{L-2-(N-1)}C_{L-N-1}^{N-1} \\	
	=&-\sum_{N=0}^{(L-1)/2}\frac{N}{(L-N-1)(L-N)}C_{L-N-1}^N+\sum_{N=0}^{(L-1)/2}\frac{L-1}{L-N-1}C_{L-N-1}^N+\\
	&\sum_{N=1}^{(L-1)/2}\frac{L-2N}{(L-N-1)(L-N)}C_{L-N-1}^{N-1} +\sum_{N=1}^{(L-1)/2}\frac{L-2}{L-2-(N-1)}C_{L-N-1}^{N-1}\\
		=&-\sum_{N=1}^{(L-1)/2}\frac{N}{(L-N-1)(L-N)}C_{L-N-1}^N+\sum_{N=0}^{(L-1)/2}\frac{L-1}{L-N-1}C_{L-N-1}^N+\\
	&\sum_{N=1}^{(L-1)/2}\frac{L-2N}{(L-N-1)(L-N)}C_{L-N-1}^{N-1} +\sum_{N=1}^{(L-1)/2}\frac{L-2}{L-2-(N-1)}C_{L-N-1}^{N-1}\\
		=& \sum_{N=0}^{(L-1)/2}\frac{L-1}{L-1-N}C_{L-1-N}^N+\sum_{N=1}^{(L-1)/2}\frac{L-2}{L-2-(N-1)}C_{L-2-(N-1)}^{N-1}\\
		=& \sum_{N=0}^{(L-1)/2}\frac{L-1}{L-1-N}C_{L-1-N}^N+\sum_{N=0}^{(L-2-1)/2}\frac{L-2}{L-2-N}C_{L-2-N}^{N}\\
			=&G_{L-1}+G_{L-2}.
	\end{align*}
For an even $L$, we have
	\begin{align*}
		G_L=&\sum_{N=0}^{L/2}\frac{L}{L-N}C_{L-N}^N =1+\sum_{N=1}^{L/2-1}\frac{L}{L-N}C_{L-N}^N+2=1+\sum_{N=1}^{L/2-1}\frac{L}{L-N}(C_{L-N-1}^N+C_{L-N-1}^{N-1})+2\\
		=&1+\sum_{N=1}^{L/2-1}\frac{L}{L-N}C_{L-N-1}^N+\sum_{N=1}^{L/2-1}\frac{L}{L-N}C_{L-N-1}^NC_{L-N-1}^{N-1}+2\\
		=&\sum_{N=0}^{L/2-1}\frac{L}{L-N}C_{L-N-1}^N+\sum_{N=1}^{L/2}\frac{L}{L-N}C_{L-N-1}^{N-1}\\
		=&\sum_{N=0}^{L/2-1}(\frac{L}{L-N}-\frac{L-1}{L-N-1})C_{L-N-1}^N+\sum_{N=0}^{L/2-1}\frac{L-1}{L-N-1}C_{L-N-1}^N+\\
		&\sum_{N=1}^{L/2}(\frac{L}{L-N}-\frac{L-2}{L-2-(N-1)})C_{L-2-(N-1)}^{N-1}+\sum_{N=1}^{L/2}\frac{L-2}{L-2-(N-1)}C_{L-2-(N-1)}^{N-1}\\
		=&-\sum_{N=0}^{L/2-1}\frac{N}{(L-N-1)(L-N)}C_{L-N-1}^N+\sum_{N=0}^{L/2-1}\frac{L-1}{L-N-1}C_{L-N-1}^N+\\
     	&\sum_{N=1}^{L/2}\frac{L-2N}{(L-N-1)(L-N)}C_{L-N-1}^{N-1}+\sum_{N=1}^{L/2}\frac{L-2}{L-2-(N-1)}C_{L-2-(N-1)}^{N-1} \\
		=&-\sum_{N=1}^{L/2-1}\frac{N}{(L-N-1)(L-N)}C_{L-N-1}^N+\sum_{N=0}^{L/2-1}\frac{L-1}{L-N-1}C_{L-N-1}^N+\\
     	&\sum_{N=1}^{L/2-1}\frac{L-2N}{(L-N-1)(L-N)}C_{L-N-1}^{N-1}+\sum_{N=1}^{L/2}\frac{L-2}{L-2-(N-1)}C_{L-2-(N-1)}^{N-1} \\
		=&\sum_{N=0}^{L/2-1}\frac{L-1}{L-1-N}C_{L-N-1}^N+\sum_{N=1}^{L/2}\frac{L-2}{L-2-(N-1)}C_{L-2-(N-1)}^{N-1}\\
		=&\sum_{N=0}^{L/2-1}\frac{L-1}{L-1-N}C_{L-1-N}^N+\sum_{N=0}^{L/2-1}\frac{L-2}{L-2-N}C_{L-2-N}^{N}
	\\	=&G_{L-1}+G_{L-2}.
	\end{align*}
\end{widetext}
Here, the relations $C_{L-N}^N=C_{L-N-1}^N+C_{L-N-1}^{N-1}$ and $C_{L-N-1}^{N}=(L-2N)/N \; C_{L-N-1}^{N-1}$ have been exploited, where $1\le N\le (L-1)/2$ for an odd $L$ and $1\le N\le L/2-1$ for an even $L$, respectively.  As a result, it is readily seen that $\{ G_L\}$ constitutes the Fibonacci sequence, given $G_1=1$ and $G_2=3$. Hence, we have
\begin{equation}
	G_{L}=(\frac{3+\sqrt{5}}{2})^{L/2} + (\frac{3-\sqrt{5}}{2})^{L/2}.
	\label{bL}
\end{equation}
Note that, for the one-dimensional flat-band ferromagnetic Tasaki model, the system size $L$ is always even. The desired result (cf. Eq.~(2)) thus follows from  the relation between ${\rm Dim }\;(\Omega_L^{\rm PBC})$ and $G_L$, when $L$ is even. 

 A similar derivation also works for the ground-state degeneracy under OBCs.  We introduce the ground-state degeneracy  under OBCs, denoted as ${\rm dim }(\Omega_L^{\rm OBC})(N)$,  in the sector with the number of electrons $N$ fixed.  Since the Hamiltonian (\ref{ham}) is pathological, in the sense that the lowest energy eigenvectors in the sectors with different $N$ are degenerate, it makes sense to define the total ground-state degeneracy ${\rm Dim }(\Omega_L^{\rm OBC})$  under OBCs as the sum over all the sectors with different eigenvalues of ${\hat N}$ that contain degenerate ground states.  That is, we have ${\rm Dim }(\Omega_L^{\rm OBC})=\sum_{N=0}^{L/2}{\rm dim }(\Omega_L^{\rm OBC})(N)$, given that degenerate ground states are absent in the sectors when $N$ is beyond $L/2$.  It is found that ${\rm dim }(\Omega_L^{\rm OBC})(N)=C_{L+1-N}^N$, consistent with the exact diagonalization results, as listed in Table~\ref{tab:gsobc}. As it turns out, ${\rm Dim }(\Omega_{L}^{\rm OBC})$ satisfies the three-term recursive relation ${\rm Dim }(\Omega_{L+4}^{\rm OBC})=3{\rm Dim }(\Omega_{L+2}^{\rm OBC})-{\rm Dim }(\Omega_{L}^{\rm OBC})$.
 This is formally identical to the three-term recursive relation for  the ground-state degeneracy in the staggered  ${\rm SU}(3)$ spin-1 ferromagnetic biquadratic model~\cite{smGolden} under OBCs. As follows from the argument there, we are able to derive the desired result (cf. Eq.~(3)). 

In particular, if $L$ is sufficiently large, then we have ${\rm Dim }\;(\Omega_L^{\rm PBC}) \sim \phi ^{L}$, where $\phi = (1+{\sqrt 5})/2$ is the golden ratio. This is also valid for ${\rm Dim }\;(\Omega_L^{\rm OBC})$. As a consequence, the ground-state degeneracies  ${\rm Dim }\;(\Omega_L^{\rm PBC})$ and  ${\rm Dim }\;(\Omega_L^{\rm OBC})$ behave asymptotically as the golden spiral - a self-similar geometric object.

\begin{table}
		\renewcommand{\arraystretch}{1.5}
	\begin{tabular}{c|ccccccc|cc}
		\hline \hline
		${\rm dim }(\Omega_L^{\rm OBC})(N)$&  &  &  &$N$ & & & &${\rm Dim }(\Omega_L^{\rm OBC})$   \\
		\hline 
		$L$\;\;\;\;\;& \;\;0\;\; & \;\;1\;\; &\;\;\; 2 \;\;\;&\;\;3\;\;& \;\;4\;\; & \;\;5\;\;  & \;\;6\;\;& \;\;\;\;\\
		\hline 
		4\;\; & 1 & 4 & 3  &  &   &  & & 8\\
		6\;\;& 1 & 6 & 10 & 4 &  & &&21\\
		8 \;\;& 1 & 8 & 21 & 20& 5& &&55\\
		10 \;\;& 1 & 10 & 36 & 56& 35& 6&&144\\
		12\;\;& 1 & 12 & 55 & 120& 126& 56&7&377\\
		\hline \hline
	\end{tabular}
	\caption{The ground-state degeneracies ${\rm Dim }(\Omega_L^{\rm OBC})$ and ${\rm dim }(\Omega_L^{\rm OBC})(N)$ from the exact diagonalization for the one-dimensional flat-band ferromagnetic Tasaki model under OBCs, where the system size $L$ is up to $L=12$. }
	\label{tab:gsobc}  
\end{table}

\subsection{An exact graded MPS representation for the orthonormal basis states}

We now turn to a detailed construction of an exact graded MPS representation for the orthonormal basis states at and away from quarter filling. Indeed,
the presence of the exact Schmidt decompositions for the orthonormal basis states arising from SSB with type-B GMs implies that they admit an exact graded MPS representation, which is adapted from the standard MPS representation for quantum  many-body spin systems.
Here, we restrict ourselves to the orthonormal basis states $|L,M\rangle$ at quarter filling, $|L,M\rangle_4$ at $1/8$ filling and $|L,M\rangle_{2,4}$ at $1/12$ filling.

\subsubsection {An exact graded MPS representation for the highest weight state $|\psi_0\rangle$ at quarter filling}

Let us start from an exact graded MPS representation for the highest weight state $|\psi_0\rangle$ for the one-dimensional flat-band ferromagnetic Tasaki model at quarter filling. 
For this purpose,  it is necessary to work in a $Z_2$-graded vector space, with the parity of each (local) orthonormal basis state being either even or odd, in order to take into account the anti-commutation relations for itinerant electrons. Throughout this work, we refer to a $Z_2$-graded vector space as a graded vector space.
Originally, the notion of the graded vector spaces has been exploited in Refs.~\cite{smgradpeps} in a graded PEPS representation for the ground state in the two-dimensional $t-J$ model as a conceptual development of the so-called fermionic PEPS representation~\cite{smkraus} (see also Refs.~\cite{smfermionpeps1,smfermionpeps2}). In the context of a graded MPS representation,   we need to introduce a graded vector space, as the local Hilbert  space, at each lattice site and an auxiliary space on each bond. Hence, it is necessary to introduce parities for both the physical indices and the bond indices,  labeling the basis states in the local Hilbert  space at each lattice site and in the auxiliary space on each bond. 
We remark that any tensor must be represented in a form, with the addition of all the parities of the indices being zero (mod 2). In addition, as a convention,  for a given tensor at a lattice  site, the basis states in the graded vector spaces should be ordered as follows. The basis states in the auxiliary graded vector space on the left hand side comes first, then  the basis states in the physical space state and finally the  basis states in the auxiliary graded vector space on the right hand side.

At each lattice site $x$ in $\Lambda$, $\alpha_x$ denotes the physical index representing the four basis states $|\alpha_x \rangle$ ($\alpha_x =1,2,3$, and 4), with
$|1_x \rangle \equiv |0\rangle_x$,  $|2_x \rangle \equiv|\uparrow\rangle_x$, $|3_x \rangle \equiv |\downarrow\rangle_x$ and $|4_x \rangle \equiv|\uparrow\downarrow\rangle_x$, with the parities $[1_x]=[4_x]=0$ and $[2_x]=[3_x]=1$.  Note that the parities $[\alpha_x]$ are the same for the dual basis states $\langle \alpha_x|$. With this fact in mind, we are now in a position to follow the prescription in Ref.~\cite{smgradpeps} to 
construct an exact graded MPS representation for the highest weight state $|\psi_0\rangle$.

First, the fermionic Fock vacuum state $|0\cdots0\rangle$ may be represented in terms of a trivial graded MPS representation, where the bond dimension $\chi$ is one, with its parity being even. Mathematically, we have $\otimes _x |1_x  \rangle$. Second,
$\hat{a}_{p,\uparrow}^\dagger$  is represented in terms of a graded MPO representation,
consisting of the tensors $W_x$, with $x=p-1/2, p$ and $p+1/2$:  
\begin{equation}
	W_{x}=(W_{x})_{l_xr_x}^{\alpha_{x} \beta_{x}}|l_x\rangle|\alpha_{x}\rangle\langle \beta_{x}|\langle r_x|,
\end{equation}
where   $\langle r_{p-1/2}|$ and $\langle r_p|$ being the basis states in the dual auxiliary space of auxiliary spaces  $|l_p\rangle$ and   $|l_{p+1/2}\rangle$, each of which represents the two auxiliary basis states, i.e., $|1_p\rangle$ and $|2_p\rangle$, and $|1_{p+1/2}\rangle$, and $|2_{p+1/2}\rangle$, with the parities $[1_p]=[1_{p+1}]=0$ and $[2_p]=[2_{p+1/2}]=0$. In addition, $|l_{p-1/2}\rangle$ and $\langle r_{p+1/2}|$ being the (dual) auxiliary space of one basis state with an even and an odd parity, i.e., $|1_{p-1/2}\rangle$ and $\langle 1_{p+1/2}|$. 
Such definitions of parities on each bond keep $(W^{x})_{l_xr_x}^{\alpha_{x} \beta_{x}}$ ($x=p-1/2, p$ and $p+1/2$) even.
As a result, the three tensors are all even that can move freely in such a tensor network representation. Note that, one has freedom to choose the parity of $|1_{p-1/2}\rangle$ to be even or odd. Accordingly, the other (dual) auxiliary space  will be changed if  $|1_{p-1/2}\rangle$ is odd.  
According to the definition of $\hat{a}_{p,\uparrow}^\dagger$, we write down the non-zero components of the three tensors:
for $x=p-1/2$, $(W_{x})_{1_x 1_x}^{\alpha_x\beta_x}=\delta_{\alpha_x,\beta_x}$,  $(W_{x})_{1_x2_x}^{1_x2_x}=-\nu$;
for $x=p$, $(W_{x})_{1_x1_x}^{\alpha_x\beta_x}= (W_{x})_{2_x2_x}^{\alpha_x\beta_x}=\delta_{\alpha_x,\beta_x}$,  $(W_{x})_{1_x2_x}^{1_x2_x}=1$;
for $x=p+1/2$, $(W_{x})_{1_x1_x}^{1_x2_x}= -\nu$,  $(W_{x})_{2_x1_x}^{\alpha_x\beta_x}=\delta_{\alpha_x,\beta_x}$.

Then, we are able to construct a graded MPS representation for the highest weight state at quarter filling by contracting the graded MPO representations for  $\hat{a}_{p}^{\dagger}$'s with the trivial graded MPS representation for the fermionic Fock vacuum state $|0\cdots 0\rangle$. 
To this end, we define the two tensors  $A$ and $B$ for the site $p$ and $u$ with $p\in\mathscr{E}$ and $u\in\mathscr{I}$.
Specifically,
\begin{equation*}
	A=\sum_{l_pr_p\alpha_p}A_{l_p r_p}^{\alpha_p}|l_p\rangle\langle \alpha_{p}|\langle r_p|,
\end{equation*}
and
\begin{equation*}
	B=\sum_{l_ur_u\alpha_u}B_{l_u r_u}^{\alpha_u}|l_u\rangle\langle \alpha_{u}|\langle r_u|,
\end{equation*}
where the non-zero components of the two tensors $A$ and $B$ are
\begin{widetext}
\begin{align*}
	A_{1_p 1_p}^{1_p}=	A_{2_p 2_p}^{1_p}=	A_{1_p 2_p}^{2_p}=1,
	B_{1_u 1_u}^{2_u}=	B_{2_u 2_u}^{2_u}=-\nu,\;	B_{2_u 1_u}^{1_u}=1.
\end{align*}
Hence, the highest weight state $|{\rm \psi}_0\rangle$ under PBCs admits an exact graded MPS representation, consisting of the two tensors $A$  and $B$:

	\begin{equation}
		|{\rm\psi}_0\rangle=\sum_{\{\alpha_x\}\;,x \in \Lambda}C(A^{\alpha_1}B^{\alpha_{3/2}}\cdots A^{\alpha_p}B^{\alpha_{p+1/2}} \cdots A^{\alpha_{L/2}}B^{\alpha_{L/2+1/2}})|\alpha_1\alpha_{3/2}\cdots\alpha_p\alpha_{p+1/2}\cdots\alpha_{L/2}\alpha_{L/2+1/2}\rangle.
		\label{psi0}	
	\end{equation}
	Here, $C$ represents a contraction operation. Specifically, we have
	\begin{equation}
		|\psi_0\rangle=\sum_{\{\alpha_x\}\;,x \in \Lambda}\sum_\beta 	{\rm sign}(\beta) (A^{\alpha_1}B^{\alpha_{3/2}}\cdots A^{\alpha_p}B^{\alpha_{p+1/2}} \cdots A^{\alpha_{L/2}}B^{\alpha_{L/2+1/2}})_{\beta\beta}|\alpha_1\alpha_{3/2}\cdots\alpha_p\alpha_{p+1/2}\cdots\alpha_{L/2}\alpha_{L/2+1/2}\rangle,
	\end{equation}
\end{widetext}
with 
\begin{equation*}
	{\rm sign}(\beta)=(-1)^{L/2+1-[L/4]}(-1)^{(L/2+1)n_{\beta}}.
\end{equation*}
That is, the contraction operation $C$ is formally identical to the so-called supertrace or trace in the auxiliary graded vector space, up to an overall minus sign that depends on $L$.
Here $[L/4]$ means the largest integer less than or equal to $L/4$. 

Hence, the normalization factor $Z_0(L)$ of the highest weight ground state results from the eigenvalues of the transfer matrix $E$. Explicitly, we have

\begin{equation} 
	Z_0(L)\!=\!\sqrt{(\lambda_{+}^e)^{L/2}\!+\!(\lambda_{-}^e)^{L/2}\!+\!(-1)^{L/2\!+\!1}[(\lambda_{1}^o)^{L/2}\!+\!(\lambda_{2}^o)^{L/2}]}.
	\label{zl}
\end{equation}	
Here $\lambda_{\pm}^e=(1+2\nu^2\pm\sqrt{1+4\nu^2})/2$ and $\lambda_{1}^o=\lambda_{2}^o=\nu^2$ are eigenvalues of the even and odd blocks in the transfer matrix $E$, respectively. 
This expression turns out to be identical to that presented in the main text.

According to the definition of $\hat{a}_{p,\uparrow}^\dagger$ under OBCs, the non-zero components of the three tensors
are different at $x=1$, which take the form $(W_{x})_{1_x1_x}^{\alpha_x\beta_x}=\delta_{\alpha_x,\beta_x}$ and  $(W_{x})_{1_x2_x}^{1_x2_x}=1$.
Hence, a graded MPS representation $|{\rm \psi}_0\rangle$ for the highest weight state under OBCs at quarter-filling may be constructed by contracting the MPO representations for $\hat{a}_{p,\uparrow}^\dagger$'s with the MPS representation for the fermionic Fock vacuum state $|0\cdots 0\rangle$, consisting of the two tensors $A$  and $B$ in the bulk and two tensors $A_{1}$ and $B_{L/2+1/2}$ at the two ends:
\begin{widetext} 
	\begin{equation}
	|{\rm\psi}_0\rangle=\sum_{\{\alpha_x\},\;x \in \Lambda}(A_{1})^{\alpha_1}B^{\alpha_{3/2}}\cdots A^{\alpha_p}B^{\alpha_{p+1/2}} \cdots A^{\alpha_{L/2}}(B_{L})^{\alpha_{L/2+1/2}}|\alpha_1\alpha_{3/2}\cdots\alpha_p\alpha_{p+1/2}\cdots\alpha_{L/2}\alpha_{L/2+1/2}\rangle,
	\label{psi0obc}
	\end{equation}
	where
	\begin{equation*}
	A_1=\sum_{l_1r_1\alpha_1}(A_1)_{l_1 r_1}^{\alpha_1}|l_1\rangle\langle \alpha_{1}|\langle r_1|,
	\end{equation*}
	and
	\begin{equation*}
	B_{L/2+1/2}=\sum_{l_{L/2+1/2}r_{L/2+1/2}\alpha_{L/2+1/2}}(B_{L/2+1/2})_{l_{L/2+1/2} r_{L/2+1/2}}^{\alpha_{L/2+1/2}}|l_{L/2+1/2}\rangle\langle \alpha_{L/2+1/2}|\langle r_{L/2+1/2}|,
	\end{equation*}
\end{widetext}
	where the non-zero components of the two tensors are
	\begin{align*}
	(A_1)_{1_1 1_1}^{1_1}&=(A_1)_{1_1 2_1}^{2_1}=1; \\
	(B_{L/2+1/2})_{1_{L/2+1/2} 1_{L/2+1/2}}^{2_{L/2+1/2}}&=-\nu,\; B_{2_{L/2+1/2} 1_{L/2+1/2}}^{1_{L/2+1/2}}=1.
	\end{align*}

As an application, we may take advantage of the exact MPS representation for the highest weight state $|\psi_0\rangle$ to construct its Schmidt decomposition  under PBCs and OBCs. As it turns out, the Schmidt rank is four under PBCs and two under OBCs, given that the bond dimension is two.
As a result, the entanglement entropy for the  the highest weight state $|\psi_0\rangle$  is 2 under PBCs and 1 under OBCs, as long as $n$ is large enough. That is, it is tightly bounded from above, as follows from the area law. This is consistent with Ref.~\cite{kensuke}. 

For simplicity, we restrict to  the Schmidt decomposition under OBCs.  It takes the form
\begin{equation}
	|\psi_0\rangle=\sum_{\alpha=1}^2 \lambda_{\alpha}|\phi_\alpha^{\cal B}\rangle\otimes|\phi_\alpha^{\cal E}\rangle,
\end{equation}
where  $\lambda_1$ and $\lambda_2$ are two Schmidt coefficients 
\begin{align*}
	\lambda_{1}&=\frac{Z_0(n)Z_0(L-n)}{Z_0(L)},\\
	\lambda_{2}&=\nu\frac{\sqrt{(Z_0(n))^2-\nu^2(Z_0(n-2))^2}Z_0(L-n-1)}{Z_0(L)},
\end{align*} 
with $Z_0(x)$
\begin{equation*} 
	Z_0(x)=\sqrt{\sum_{r=0}^{x/2}C_{x-r}^r\nu^{2r}},
\end{equation*}
and $|\phi_\alpha^{\cal B}\rangle$ and $|\phi_\alpha^{\cal E}\rangle$ are the Schmidt basis states for the block and the environment, respectively.
They take the form
\begin{align*}
	|\phi_1^{\cal B}\rangle&= \frac {1} {Z_0(n)} \; \prod_{p\in [1,2,\ldots, n/2]} \hat{a}_{p,\uparrow}^\dagger |0\cdots 0 \rangle,  \\
	|\phi_2^{\cal B}\rangle&= -\frac {1} {\sqrt{(Z_0(n))^2-\nu^2(Z_0(n-2))^2}} \;  \times \\
	&\prod_{p\in[1,2,\ldots, n/2]} \hat{a}_{p,\uparrow}^\dagger \hat{c}^\dagger_{n/2+1/2,\uparrow}|0\cdots 0 \rangle,  
\end{align*}
and
\begin{align*}
	|\phi_1^{\cal E}\rangle&= \frac {1} {Z_0(L-n)}\; \prod_{p\in [n/2+1,\ldots, L/2]} \hat{a}_{p,\uparrow}^\dagger|0\cdots 0 \rangle,  \\
	|\phi_2^{\cal E}\rangle&= \frac {1}{Z_0(L-n-1)} \;  \times \\
	&\prod_{p\in [n/2+2,\ldots, L/2]} \hat{a}_{p,\uparrow}^\dagger|0\cdots 0 \rangle.  
\end{align*}
Here $\hat{a}_{1,\uparrow}^\dagger$ and $\hat{a}_{n/2+1,\uparrow}^\dagger$ are defined as ${\hat a}_{1,\uparrow}^\dagger={\hat c}_{1,\uparrow}^\dagger-\nu{\hat c}_{3/2,\uparrow}^\dagger$ and ${\hat a}_{n/2+1,\uparrow}^\dagger={\hat c}_{n/2+1,\uparrow}^\dagger-\nu{\hat c}_{n/2+3/2,\uparrow}^\dagger$.

We stress that the number of electrons $N$ for $|\phi_1^{\cal B}\rangle$ and $|\phi_2^{\cal B}\rangle$ are $n/2$ and $n/2+1$, and  the number of electrons $N$ for $|\phi_1^{\cal E}\rangle$ and $|\phi_2^{\cal E}\rangle$ are $(L-n)/2$ and $(L-n)/2-1$, respectively. As a result, the fillings for the block and the environment are away from quarter filling for the highest weight state $|\psi_0\rangle$ at quarter filling.
As a consequence, the block $\cal B$ and the environment $\cal E$, as a subsystem, do not share the same type of the orthonormal basis states as the entire system.

\subsubsection {An exact graded MPS representation for the highest weight state $|\psi_0\rangle_q$ at $1/8$ filling}

At $1/8$ filling, it is straightforward to construct an exact graded MPS representation for the highest weight state $|\psi_0\rangle_q$ with $q=4$ by  contracting the graded MPO representations for $\hat{a}_{p,\uparrow}^\dagger$'s with the trivial graded MPS representation for the fermionic Fock vacuum state $|0\cdots 0\rangle$,  which consists of the four tensors  $A$, $B$, $C$ and  $D$,
\begin{widetext}
	\begin{equation*}
		|\psi_0\rangle=\sum_{\{\alpha_x\}\;,x \in \Lambda}(A^{\alpha_1}\cdots A^{\alpha_p}B^{\alpha_{p+1/2}} C^{\alpha_{p+1}}D^{\alpha_{p+3/2}}\cdots D^{\alpha_{L/2+1/2}})|\alpha_1\alpha_{3/2}\cdots\alpha_p\alpha_{p+1/2}\alpha_{p+1}\alpha_{p+3/2}\cdots\alpha_{L/2+1/2}\rangle.	\end{equation*}
\end{widetext}
Specifically, at the lattice sites $p=1,\;3,\cdots,L/2-2$, we have
\begin{equation*}
	A=\sum_{l_pr_p\alpha_p}A_{l_p r_p}^{\alpha_p}|l_p\rangle\langle \alpha_{p}|\langle r_p|,
\end{equation*}
with $l_p=r_p=1$;
at the lattice sites $u=3/2,\;7/2,\cdots,L/2-1/2$, we have
\begin{equation*}
	B=\sum_{l_ur_u\alpha_u}B_{l_u r_u}^{\alpha_u}|l_u\rangle\langle \alpha_{u}|\langle r_u|,
\end{equation*}
with $l_u=1$, and $r_u=1$ and 2;
at the sites $p=2,\;4,\cdots,L/2$, we have
\begin{equation*}
	C=\sum_{l_pr_p\alpha_p}C_{l_p r_p}^{\alpha_p}|l_p\rangle\langle \alpha_{p}|\langle r_p|,
\end{equation*}
with $l_p$ and $r_p$ taking two values $1$ and $2$;
at the sites $u=5/2,\;9/2,\cdots,L/2+1/2$, we have
\begin{equation*}
	D=\sum_{l_ur_u\alpha_u}D_{l_u r_u}^{\alpha_u}|l_u\rangle\langle \alpha_{u}|\langle r_u|,
\end{equation*}
with $l_u=1$ and $2$, and $r_u=1$, 
where the non-zero components of the four tensors are
\begin{align*}
	A_{1_p 1_p}^{1}&=1; \\
	B_{1_u 1_u}^{1_u}&=1,\;	B_{1_u 2_u}^{2_u}=-\nu; \\
	C_{1_p 1_p}^{1_p}&=	C_{2_p 2_p}^{1_p}=	C_{1_p 2_p}^{2_p}=1; \\
	D_{1_u 1_u}^{2_u}&=-\nu,\;	D_{2_u 1_u}^{1_u}=1.
\end{align*}

\begin{widetext} 
\subsubsection {An exact graded MPS representation for the highest weight state $|\psi_0\rangle_{q_1,q_2}$ at $1/12$ filling}

At $1/12$ filling, it is straightforward to construct an exact graded MPS representation for the highest weight state $|\psi_0\rangle_{q_1,q_2}$ with  $q_1=2$ and $q_2=4$ by contracting the graded MPO representations for $\hat{a}_{p,\uparrow}^\dagger$'s with the trivial graded MPS representation for the fermionic Fock vacuum state $|0\cdots 0\rangle$, which consists of the four tensors $A$, $B$, $C$ and $D$,

	\begin{equation*}
	|\psi_0\rangle_{2,4}=\sum_{\{\alpha_x\}\;,x \in \Lambda}(A^{\alpha_1}\cdots A^{\alpha_p}A^{\alpha_{p+1/2}} A^{\alpha_{p+1}}B^{\alpha_{p+3/2}}C^{\alpha_{p+2}}D^{\alpha_{p+5/2}}\cdots D^{\alpha_{L/2+1/2}})|\alpha_1\alpha_{3/2}\cdots\alpha_p\alpha_{p+1/2}\alpha_{p+1}\alpha_{p+3/2}\alpha_{p+2}\alpha_{p+5/2}\cdots\alpha_{L/2+1/2}\rangle.
	\end{equation*}
\end{widetext}
Specifically, at the sites $p=1,\;4,\cdots,L/2-2$, we have
\begin{equation*}
A=\sum_{l_pr_p\alpha_p}A_{l_p r_p}^{\alpha_p}|l_p\rangle\langle \alpha_{p}|\langle r_p|,
\end{equation*}
with $l_p=r_p=1$; at the sites $u=3/2,\;9/2,\cdots,L/2-3/2$, we have
\begin{equation*}
A=\sum_{l_ur_u\alpha_u}A_{l_u r_u}^{\alpha_u}|l_u\rangle\langle \alpha_{u}|\langle r_u|,
\end{equation*}
with $l_u=r_u=1$; at the sites $p=2,\;5,\cdots,L/2-1$, we have
\begin{equation*}
A=\sum_{l_pr_p\alpha_p}A_{l_p r_p}^{\alpha_p}|l_p\rangle\langle \alpha_{p}|\langle r_p|,
\end{equation*}
with $l_p=r_p=1$;  at the sites $u=5/2,\;11/2,\cdots,L/2-1/2$, we have
\begin{equation*}
B=\sum_{l_ur_u\alpha_u}B_{l_u r_u}^{\alpha_u}|l_u\rangle\langle \alpha_{u}|\langle r_u|,
\end{equation*}
with $l_u=1$, and $r_u=1$ and 2;
at the sites $p=3,\;6,\cdots,L/2$, we have
\begin{equation*}
C=\sum_{l_pr_p\alpha_p}C_{l_p r_p}^{\alpha_p}|l_p\rangle\langle \alpha_{p}|\langle r_p|,
\end{equation*}
with $l_p$ and $r_p$ taking two values $1$ and $2$;
at the sites $u=7/2,\;13/2,\cdots,L/2+1/2$, we have 
\begin{equation*}
D=\sum_{l_ur_u\alpha_u}D_{l_u r_u}^{\alpha_u}|l_u\rangle\langle \alpha_{u}|\langle r_u|,
\end{equation*}
with $l_u=1$ and $2$, and $r_u=1$, 
where the non-zero components of the four tensors are
\begin{align*}
A_{1_p 1_p}^{1}&=A_{1_u 1_u}^{1}=1;\nonumber \\
B_{1_u 1_u}^{1_u}&=1,\;	B_{1_u 2_u}^{2_u}=-\nu;\nonumber \\
C_{1_p 1_p}^{1_p}&=	C_{2_p 2_p}^{1_p}=	C_{1_p 2_p}^{2_p}=1;	\nonumber \\
D_{1_u 1_u}^{2_u}&=-\nu,\;	D_{2_u 1_u}^{1_u}=1.
\end{align*}

\subsubsection{A graded MPO representation of the power of the lowering operator $({\hat S}^-)^M$}

A power of the lowering operator $({\hat S}^-)^M$ is always even, and only acts on the two (local) singly-occupied basis states.
As it turns out, a graded MPO representation for $({\hat S}^-)^M$ is formally identical to the counterpart for the $\rm {SU(2)}$ quantum spin many-body  systems~\cite{smexactmps}, which is adapted from that for a generic operator, developed in Ref.\;\cite{smyuping}, to the lowering operator $S^-$ of the symmetry group ${\rm SU}(2)$. Hence, we have

\begin{equation}
	({\hat S}^-)^M=W_{[1]}\cdots W_{[x]}\cdots W_{[L/2+1/2]} \,,
	\label{lo1}
\end{equation}
where\begin{equation*}
	W_{[x]}=\sum_{l_xr_x\alpha_x\beta_x}(W_{[x]})_{l_xr_x}^{\alpha_x\beta_x}|l_x\rangle|\beta_x\rangle\langle \alpha_x|\langle r_x|,
\end{equation*}
with $l_x=1$ at $x=1$, $r_x=1$ at $x=L/2+1/2$, and $l_x$ ($1<x\le L/2+1/2$) and  $r_x$ ($1\le x<L/2+1/2$) range from $1$ to $M+1$.  
Here the non-zero components of the two tensors $W_{[1]}$ and $W_{[L/2+1/2]}$ at the two ends are
\begin{align*}
	&(W_{[1]})_{1_1,M_1}^{2_13_1}=M,	(W_{[1]})_{1_1 (M+1)_1}^{\alpha_1\beta_1}=\delta_{\alpha_1,\beta_1};\nonumber \\
	&(W_{[L/2+1/2]})_{1_{L/2+1/2} 1_{L/2+1/2}}^{\alpha_{L/2+1/2} \beta_{L/2+1/2}}=\delta_{\alpha_{L/2+1/2},\beta_{L/2+1/2}},		(W_{[L/2+1/2]})_{2_{L/2+1/2} 1_{L/2+1/2}}^{2_{L/2+1/2} 3_{L/2+1/2}}=1,
\end{align*}
and the non-zero components of the tensors $W_{[x]}$ ($x=3/2,\;\ldots,\; L/2$) in the bulk are
\begin{align*}
	(W_{[x]})_{l_x l_x}^{\alpha_x \beta_x}=\delta_{\alpha_x,\beta_x},	(W_{[x]})_{l_x (l-1)_x}^{2_x3_x}=1\; {\rm for }\; 2\le l \le M+1. 
\end{align*}

\subsubsection{An exact graded MPS representation for  the orthonormal basis states}

An exact graded MPS representation for the orthonormal basis states $|L,M\rangle=1/Z(L,M)({\hat S}^-)^M|\psi_0\rangle$ under PBCs follows from contracting the graded MPO representation for the power of the lowering operator $({\hat S}^-)^M$ with the graded MPS representation for the highest weight state $|\psi_0\rangle$.
Hence, the highly degenerate ground states $|L,M\rangle$ admit an exact graded MPS representation,
\begin{widetext}
	\begin{equation}
		|L,M\rangle = \sum_{\{\alpha_x\},\;x\in \Lambda}  C(X^{\alpha_1}F^{\alpha_{3/2}}\cdots E^{\alpha_{p}}F^{\alpha_{p+1/2}}\cdots E^{\alpha_{L/2}}Y^{\alpha_{L/2+1/2}})|\alpha_1\alpha_{3/2}\cdots\alpha_{p}\alpha_{p+1/2}\cdots\alpha_{L/2}\alpha_{L/2+1/2}\rangle,
		\label{smlmDEFG}
	\end{equation}
\end{widetext}
where the two tensors $X^{\alpha_1}$ and $Y^{\alpha_{L/2+1/2}}$ at the two ends take the form
\begin{equation*}
	X^{\alpha_1}=\sum_{\upsilon}
	(W_{[1]})^{\alpha_1\upsilon}A^{\upsilon},
\end{equation*}
\begin{equation*}
	Y^{\alpha_{L/2+1/2}}=\sum_{\upsilon}
	(W_{[L/2+1/2]})^{\alpha_{L/2+1/2}\upsilon}B^{\upsilon},
\end{equation*}
and the tensors $E^{\alpha_{p}}$ in the bulk with $p=2,\;3,\;\ldots, \;L/2$ take the form
\begin{equation*}
	E^{\alpha_{p}}=\sum_{\upsilon}
	(W_{[p]})^{\alpha_{p}\upsilon}A^{\upsilon},
\end{equation*}
and the  tensors $F^{\alpha_{u}}$ in the bulk with $u=3/2,\;5/2,\; \ldots, \;L/2-1/2$, take the form
\begin{equation*}
	F^{\alpha_{u}}=\sum_{\upsilon}
	(W_{[u]})^{\alpha_{u}\upsilon}B^{\upsilon}.
\end{equation*}

We are thus led to the explicit expressions for an exact graded MPS representation, presented in the main text (cf. Eq.(\ref{lmDEFG}) and below). In addition, the contraction $C$ works in the same way as it did in Eq.~(\ref{psi0}), which is formally identical to the so-called supertrace or trace in the auxiliary graded vector spaces, up to an overall minus sign, depending on the values of the system size $L$.

Under OBCs, as specified in Eq.~(\ref{lmDEFGobc}) in the main text, the highly degenerate ground states $|L,M\rangle$ admit an exact graded MPS representation,
\begin{widetext}
	\begin{equation}
	|L,M\rangle = \sum_{\{\alpha_x\},\;x\in \Lambda}  (U^{\alpha_1}F^{\alpha_{3/2}}\cdots E^{\alpha_{p}}F^{\alpha_{p+1/2}}\cdots E^{\alpha_{L/2}}V^{\alpha_{L/2+1/2}})|\alpha_1\alpha_{3/2}\cdots\alpha_{p}\alpha_{p+1/2}\cdots\alpha_{L/2}\alpha_{L/2+1/2}\rangle,
	\label{smlmDEFGobc}
	\end{equation}
\end{widetext}
where the two tensors $U^{\alpha_1}$ and $V^{\alpha_{L/2+1/2}}$ at the two ends take the form
\begin{equation*}
U^{\alpha_1}=\sum_{\upsilon}
(W_{[1]})^{\alpha_1\upsilon}(A_1)^{\upsilon},
\end{equation*}
\begin{equation*}
V^{\alpha_{L/2+1/2}}=\sum_{\upsilon}
(W_{[L/2+1/2]})^{\alpha_{L/2+1/2}\upsilon}(B_{L+1/2})^{\upsilon},
\end{equation*}
and the tensors $E^{\alpha_{p}}$ in the bulk with $p=2,\;3,\;\ldots, \;L/2$ take the form
\begin{equation*}
E^{\alpha_{p}}=\sum_{\upsilon}
(W_{[p]})^{\alpha_{p}\upsilon}A^{\upsilon},
\end{equation*}
and the  tensors $F^{\alpha_{u}}$ in the bulk with $u=3/2,\;5/2,\; \ldots, \;L/2-1/2$, take the form
\begin{equation*}
F^{\alpha_{u}}=\sum_{\upsilon}
(W_{[u]})^{\alpha_{u}\upsilon}B^{\upsilon}.
\end{equation*}

It is straightforward to extend this construction to the orthonormal basis states $|L,M\rangle_q$ at a specific filling, with the filling fraction $f=1/2 -1/q$. As an example, we consider $q=4$, with the filling fraction $f$ being $1/4$. 
Given the graded MPO representation for $({\hat S}^-)^M$, an exact graded MPS representation for $({\hat S}^-)^M|\psi_0\rangle_4$ simply follows from contracting the graded MPO representations for the power of the lowering operator with an exact graded MPS representation for $|\psi_0\rangle_4$. 
As a result, we are led to an exact graded MPS representation for the orthonormal basis states $|L,M\rangle_4$. As specified in Eq.~(\ref{lm4DEFGH}) in the main text, $|L,M\rangle_4$ at $1/8$ filling takes the form 
\begin{widetext}
	\begin{equation}
		|L,M\rangle_4 = \sum_{\{\alpha_x\},\;x\in \Lambda}  (X^{\alpha_{1}}\cdots E^{\alpha_{p}}F^{\alpha_{p+1/2}}G^{\alpha_{p+1}}H^{\alpha_{p+3/2}}\cdots Y^{\alpha_{L/2+1/2}})|\alpha_1\cdots\alpha_{p} \alpha_{p+1/2}\alpha_{p+1} \alpha_{p+3/2}\cdots \alpha_{L/2+1/2}\rangle,
		\label{smlm4DEFGH}
	\end{equation}
\end{widetext}
where the two tensors  $X^{\alpha_1}$ and $Y^{\alpha_{L/2+1/2}}$ at the two ends take the form
\begin{equation*}
	X^{\alpha_1}=\sum_{\upsilon}
	(W_{[1]})^{\alpha_1\upsilon}A^{\upsilon},
\end{equation*}
\begin{equation*}
	Y^{\alpha_{L/2+1/2}}=\sum_{\upsilon}
	(W_{[{L/2+1/2}]})^{\alpha_{L/2+1/2}\upsilon}D^{\upsilon},
\end{equation*}
and the tensors  $E^{\alpha_{p}}$ ($p=3,\;5,\;\ldots, \;L/2-1$),  $F^{\alpha_{u}}$ ($u=3/2,\;7/2,\;\ldots, \;L/2-1/2$),  $G^{\alpha_{p}}$($p=2,\;4,\;\ldots, \;L/2$) and $H^{\alpha_{u}}$ ($u=5/2,\;9/2,\;\ldots, \;L/2-3/2$) in the bulk take the forms
\begin{equation*}
	E^{\alpha_{p}}=\sum_{\upsilon}
	(W_{[p]})^{\alpha_{p}\upsilon}A^{\upsilon},
\end{equation*}
\begin{equation*}
	F^{\alpha_{u}}=\sum_{\upsilon}
	(W_{[u]})^{\alpha_{u}\upsilon}B^{\upsilon},
\end{equation*}
\begin{equation*}
	G^{\alpha_{p}}=\sum_{\upsilon}
	(W_{[p]})^{\alpha_{p}\upsilon}C^{\upsilon},
\end{equation*}
and 
\begin{equation*}
	H^{\alpha_{u}}=\sum_{\upsilon}
	(W_{[u]})^{\alpha_{u}\upsilon}D^{\upsilon}.
\end{equation*}

We turn to the orthonormal basis states $|L,M\rangle_{q_1,q_2}$ at a specific filling, with the filling fraction $f=q_1/(2(q_1+q_2))$.  As an example, we consider $q_1=2$ and $q_2=4$, with the filling fraction $f$ being $1/6$. 
Given the graded MPO representation for $({\hat S}^-)^M$, an exact graded MPS representation for $({\hat S}^-)^M|\psi_0\rangle_{2,4}$ simply follows from contracting the graded MPO representations for the power of the lowering operator with an exact graded MPS representation for $|\psi_0\rangle_{2,4}$. 
As a result, we are led to an exact graded MPS representation for the orthonormal basis states $|L,M\rangle_{2,4}$. As specified in Eq.~(\ref{lm24DEFGH}) in the main text, $|L,M\rangle_{2,4}$ at $1/12$ filling takes the form 
	\begin{widetext}
	\begin{align}
 |L,M\rangle_{2,4} = \sum_{\alpha_1,\ldots,  \alpha_{p}, \alpha_{p+1/2}, \alpha_{p+1}, \alpha_{p+3/2},\atop \alpha_{p+2}, \alpha_{p+5/2},\ldots, \alpha_{L/2+1/2}}  (X^{\alpha_{1}}\cdots E^{\alpha_{p}}E^{\alpha_{p+1/2}}E^{\alpha_{p+1}}F^{\alpha_{p+3/2}}G^{\alpha_{p+2}}H^{\alpha_{p+5/2}}\cdots Y^{\alpha_{L/2+1/2}})\nonumber \\
 |\alpha_1\cdots\alpha_{p} \alpha_{p+1/2}\alpha_{p+1} \alpha_{p+3/2}\alpha_{p+2} \alpha_{p+5/2}\cdots \alpha_{L/2+1/2}\rangle,
 \end{align}
	\end{widetext}
	where the two tensors $X^{\alpha_1}$ and $Y^{\alpha_{L/2+1/2}}$ at the two ends take the form
	\begin{equation*}
	X^{\alpha_1}=\sum_{\upsilon}
	(W_{[1]})^{\alpha_1\upsilon}A^{\upsilon},
	\end{equation*}
	\begin{equation*}
	Y^{\alpha_{L/2+1/2}}=\sum_{\upsilon}
	(W_{[{L/2+1/2}]})^{\alpha_{L/2+1/2}\upsilon}D^{\upsilon},
	\end{equation*}
	and the tensors  $E^{\alpha_{p}}$ ($p=2,\;4,\;5,\;7\;8,\;\ldots, \;L/2-2$) and  $E^{\alpha_{u}}$ ($u=3/2,\;9/2,\;\ldots, \;L/2-3/2$),  $F^{\alpha_{u}}$ ($u=5/2,\;11/2,\;\ldots, \;L/2-1/2$),  $G^{\alpha_{p}}$($p=3,\;6,\;\ldots, \;L/2$) and $H^{\alpha_{u}}$ ($u=7/2,\;13/2,\;\ldots,\;L/2-5/2$) in the bulk take the forms
	\begin{equation*}
	E^{\alpha_{p}}=\sum_{\upsilon}
	(W_{[p]})^{\alpha_{p}\upsilon}A^{\upsilon},
	\end{equation*}
	\begin{equation*}
	E^{\alpha_{u}}=\sum_{\upsilon}
	(W_{[u]})^{\alpha_{u}\upsilon}A^{\upsilon},
	\end{equation*}
	\begin{equation*}
	F^{\alpha_{u}}=\sum_{\upsilon}
	(W_{[u]})^{\alpha_{u}\upsilon}B^{\upsilon},
	\end{equation*}
	\begin{equation*}
	G^{\alpha_{p}}=\sum_{\upsilon}
	(W_{[p]})^{\alpha_{p}\upsilon}C^{\upsilon},
	\end{equation*}
	and 
	\begin{equation*}
	H^{\alpha_{u}}=\sum_{\upsilon}
	(W_{[u]})^{\alpha_{u}\upsilon}D^{\upsilon}.
	\end{equation*}

\end{document}